\documentclass[12pt]{article}
\usepackage[utf8]{inputenc}
\usepackage{amsfonts}
\usepackage{latexsym}
\usepackage{amsmath,amssymb}
\usepackage{graphicx}
\usepackage{verbatim}
\usepackage{setspace}
\usepackage{physics}
\usepackage{color}
\usepackage{mathdots}
\usepackage{cancel}
\usepackage{color}
\usepackage{mathrsfs}  
\usepackage{array}
\usepackage{multirow}
\usepackage{amssymb}
\usepackage{tabularx}
\usepackage{hyperref}
\usepackage{booktabs}
\numberwithin{figure}{section}

\usepackage[letterpaper,hmargin=0.8in,vmargin=1in]{geometry}
\numberwithin{equation}{section}

\begin{document}
\unitlength = 1mm
\ \\
\vskip1cm
\begin{center}

{ \LARGE {\textsc{Soft Algebras via Bulk Double Soft Limits}}}

\vspace{0.8cm}
Sruthi A. Narayanan$^\dagger$

\vspace{1cm}

{\it ${}^\dagger$Perimeter Institute for Theoretical Physics,\\
31 Caroline Street North, Waterloo, ON N2L 2Y5, Canada }

\begin{abstract}
Soft and collinear limits of ``celestial amplitudes" give rise to infinite dimensional symmetry algebras for two-dimensional (2D) ``celestial" conformal field theories (CFTs). A small subset of these operators generates the action of the entire set recursively via the commutation relations. Insertion of this subset into celestial CFT correlators gives the boundary version of the first three terms in a soft expansion of the corresponding 4D bulk gravitational scattering amplitude. In this paper, we find that a bulk analog of this, which would allow the entire soft expansion of a gravitational amplitude to be determined via just the first three terms, does not follow trivially from the celestial algebras. We show how the interplay of soft and collinear limits results in subtleties for bulk amplitudes that do not show up in the boundary description. We also briefly comment on this construction in gauge theory.
 \end{abstract}
\vspace{0.5cm}

\vspace{1.0cm}

\end{center}

\pagestyle{empty}
\pagestyle{plain}
\newpage
\tableofcontents

\pagenumbering{arabic}

\section{Introduction}
Soft theorems in four-dimensional (4D) quantum field theory tell us how scattering amplitudes behave when we take one or more of the external particles to be energetically soft.\cite{Weinberg:1965nx} Practically, one starts with the full expression for a scattering amplitude and performs an expansion in the soft energy which results in an infinite series where each term is referred to as an ``order" in the soft expansion. In gauge and gravitational theories, it has been shown that at the lowest few orders, these terms can be written as differential operators acting on a lower point amplitude. This has proven to be crucial when trying to understand the infrared (IR) behavior of these theories.\cite{Strominger:2017zoo}

For a little over a decade, the ``celestial holography" program has used these IR properties in gauge and gravity theories to understand aspects of a co-dimension two ``celestial" conformal field theory (CFT) that lives on the celestial sphere at the boundary of spacetime. See~\cite{Pasterski:2021rjz, Raclariu:2021zjz} and references therein. One of the primary goals of this program, given what we have learned about celestial CFTs thus far, is to use two-dimensional (2D) CFT techniques to discern the implications for a bulk theory of quantum gravity.\cite{Pasterski:2016qvg} This particular goal has proven to be a rather difficult task but to date there has been a lot of progress using a variety of approaches, including twistors in the context of self-dual bulk theories~\cite{Costello:2022wso}, constructing explicit dualities between sectors of bulk and boundary theories~\cite{Ren:2022sws,Stieberger:2022zyk, Melton:2023lnz}, understanding the space of all possible celestial CFTs~\cite{Kapec:2022axw, Narayanan:2024qgb, Imseis:2025awd} and much more.

This duality relies on the fact that Mellin transforms of bulk scattering amplitudes behave like conformal correlation functions where the energy label $\omega$ has been replaced by a conformal dimension $\Delta$. These are often referred to as ``celestial amplitudes".\cite{Pasterski:2016qvg, Stieberger:2018onx, Pasterski:2021rjz} Via this mapping, bulk soft theorems can be recast as insertions of associated soft operators in the celestial CFT and these are referred to as conformally soft theorems on the boundary.\cite{Pate:2019mfs, Puhm:2019zbl} One can derive the conformally soft theorems by Mellin transforming the bulk amplitude and then carefully taking the conformal dimension of one of the operators to correspond to the particular order in the soft expansion.

Furthermore, it was shown that by transforming collinear limits of bulk scattering amplitudes to the boundary, one can get the operator product expansions (OPEs) of soft operators.\cite{Fan:2019emx, Pate:2019lpp} Then, via a standard mode expansion/contour integral procedure, one obtains algebras of these boundary operators.\cite{Guevara:2021abz} These algebras contain operators that are supposed to correspond to every single order in the soft expansion, given that they are labeled by integer values of the conformal dimension, and are thereby referred to as ``infinite dimensional symmetry algebras". For example, for bulk gravitational amplitudes, the set of boundary soft operators has integer conformal dimensions $\Delta<2$ where $\Delta=1,0,-1$ correspond to the first three terms in the soft expansion. A natural next step is to use the infinite symmetries to put constraints on the putative dual theory of gravity. To this end there were many studies of the possible allowed deformations of these algebras that were sourced by some additional bulk physics.\cite{Mago:2021wje, Ball:2021tmb, Melton:2022fsf, Ball:2022jac, Alday:2026rso} 

The power of this algebra comes from the fact that given the first three operators, the commutators recursively give the action of the entire infinite tower. However, this is \textbf{mainly a boundary level statement} since every operator contained in the algebra has not been matched to an explicit statement of bulk physics. Alternatively, we can attribute this to a clever use of conformal symmetry as was done in~\cite{Himwich:2023njb}. It is therefore natural to ask whether a similar relation exists explicitly for the bulk amplitudes. This would be incredibly useful since, for gravity, the first three terms in the soft expansion can be written as a differential operator acting on the lower point amplitude but beyond sub-subleading order, we know such a factorization does not occur, except for certain special cases like the maximum helicity violating (MHV) sector. In fact, in~\cite{Li:2018gnc} the authors have the following general expression for the soft expansion\footnote{Usually this is written already contracted with polarization vectors so that the $\mu\nu$ indices are not present.} 
\begin{eqnarray}
\mathcal{M}_{n+1,(\ell)}^{\mu\nu} & = & \sum_{i=1}^n\frac{q_\alpha q_\beta}{k_i\cdot q}\left[\frac{J_i^{\mu\alpha}J_i^{\nu\beta}}{(\ell+1)!} + \frac{1}{2}\frac{J_i^{\mu\alpha}U_i^{\nu\beta} + U_i^{\mu\alpha}J_i^{\nu\beta}}{\ell !} + \frac{1}{2}\frac{U_i^{\mu\alpha}U_i^{\nu\beta}}{(\ell-1)!}\right](q\cdot\partial_i)^{\ell-1}M_n \cr
& + & q_{\alpha_1}\cdots q_{\alpha_\ell}\sum_{i,j}L^{\mu\alpha_i}L^{\nu\alpha_j}D_\ell^{\alpha_1\cdots\tilde{\alpha_i}\cdots\tilde{\alpha}_j\cdots \alpha_\ell}
\end{eqnarray}
where $J_i, U_i$ are the spin and orbital parts of the full angular momentum operator and $\ell\geq 1$ is the order of the soft expansion ($\ell=1$ is the sub-subleading). They show, via a Ward identity argument, that beyond sub-subleading order, the soft expansion can be written as one term that is a differential operator on the lower point amplitude plus another term whose structure is generally unknown. A nice bulk realization of the boundary algebra could either help us determine more aspects of this unknown second term, or consistently generate the first term at every order. Ideally, we would like to show either of these via an actual bulk computation. This would also make contact with the bulk statements made in~\cite{Hamada:2018vrw}. However, what we find here is that even within the first three soft theorems, there are subtleties that arise due to the order of soft and collinear limits that make such an identification much more non-trivial.

This paper is organized as follows. We begin with the necessary background in soft and collinear properties of 4D amplitudes and celestial amplitudes in section~\ref{sec:motivation} to provide the motivation for the question we will try and answer in this work. We find that discussing OPEs and algebras requires a definition of a double soft limit which we discuss and define in section~\ref{sec:doublesoft}. In section~\ref{sec:usingdouble} we provide a schematic argument for why some na\"ive guesses do not work as suggested. We then look at the explicit expressions for these double soft factors in section~\ref{sec:explicitsoft} and attempt to come up with a prescription that will match the boundary results. We find that looking at the full double soft expansion, which combines different pairs of individual soft factors at each order in $\epsilon$ gives us an answer that is consistent but only up to sub-subleading order. In section~\ref{sec:commutators} we consider various commutators that are constructed from the aforementioned double soft factors and, once again, are unable to find a consistent way to match the boundary expectations. We then discuss this construction for gauge theory in section~\ref{sec:gauge} and comment on its implications for $\mathcal{N}=4$ SYM. We conclude in section~\ref{sec:discussion} with speculations on the implications of these results as well as some comments and future directions. In appendix~\ref{app:derivatives}
we have explicit expressions for arbitary derivatives of soft factors that appear in the text and in appendix~\ref{app:doublesoft} we calculate the full expressions for all the double soft factors as well as their collinear limits.
\section{Motivation}\label{sec:motivation}

In order to present our arguments as clearly as possible, we need to review some relevant background and discuss the origin of the statements that motivate this work. 
\paragraph{Soft theorems}
In quantum field theories, we can study the infrared behavior of scattering amplitudes by studying their limits as one or more particles are taken to be energetically soft. For an $n$-particle amplitude $\mathcal{M}_n(k_1,\cdots,k_n)$ we take $k_1\rightarrow \epsilon q_1$ which allows the amplitude to be written in terms of the small parameter $\epsilon$ and then we can expand in $\epsilon$.\footnote{In this work we will use $\epsilon$ as the expansion parameter while some of the literature uses the energy $\omega$ instead. The two notations are interchangeable.} In gravity, this expansion takes the following form\footnote{Similar results also exist in gauge theories but we want to concentrate on gravity in this paper.} 
\begin{equation}
\lim_{\epsilon_1\rightarrow 0}\mathcal{M}_{n+1}(\epsilon_1q_1,\cdots,k_{n+1}) = \left(\frac{1}{\epsilon}S^{(0)}_1+S^{(1)}_1 + \epsilon S^{(2)}_1\right)\mathcal{M}_n(k_2,\cdots,k_{n+1})+\mathcal{O}(\epsilon^2).
\end{equation}
The first three soft factors~\cite{Weinberg:1965nx, Cachazo:2014fwa} are given by 
\begin{equation}\label{eq:softfactors}
S_1^{(0)} = \sum_{a=2}^{n+1}\frac{(\varepsilon_1\cdot k_a)^2}{q_1\cdot k_a}, \ \ S_1^{(1)} = -i\sum_{a=2}^{n+1}\frac{(\varepsilon_1\cdot k_a)\varepsilon_{1,\nu}k_{a,\mu}J_a^{\mu\nu}}{q_1\cdot k_a}, \ \ S_1^{(2)} = -\frac{1}{2}\sum_{a=2}^{n+1}\frac{(\varepsilon_{1,\nu}k_{a,\mu}J_a^{\mu\nu})^2}{q_1\cdot k_a}
\end{equation}
where $J_a^{\mu\nu}$ is the full angular momentum operator
\begin{equation}
J_a^{\mu\nu} = i(k_a^\mu \partial_a^\nu - k_a^\nu\partial_a^\mu)+i(\varepsilon_a^\mu\tilde{\partial}_a^\nu - \varepsilon_a^\nu \tilde{\partial}_a^\mu).
\end{equation}
This contains both the spin and orbital angular momentum contributions and we have used the notation $\tilde{\partial} = \frac{\partial}{\partial\varepsilon}$ as a shorthand. The convention here gives the following commutator 
\begin{equation}
\left[J^{\mu\nu},J^{\rho\sigma}\right] = i\left(\eta^{\mu\rho}J^{\nu\sigma}-\eta^{\mu\sigma}J^{\nu\rho}-\eta^{\nu\rho}J^{\mu\sigma}+\eta^{\nu\sigma}J^{\mu\rho}\right).
\end{equation}
It is also important to note that, in general for gravitational amplitudes, the first three soft factors obey nice factorization properties but beyond that order there is only a nice factorization if we are in the MHV sector.~\cite{Hamada:2018vrw, Li:2018gnc, Guevara:2019ypd, Guevara:2025tsm} We will comment on this further later.

\paragraph{Collinear limits} In addition to soft limits, we can look at the collinear limits of amplitudes which also obey nice properties. For an amplitude $\mathcal{M}_n(k_1,\cdots, k_n)$ we let $k_1 = tP$ and $k_2=(1-t)P$ such that $k_1+k_2=P$ and $P$ is a null momentum.\cite{Dixon:2013uaa} This allows for the following expansion
\begin{equation}
\lim_{1||2}\mathcal{M}_n(k_1,\cdots,k_n) = \mbox{Split}(t,k_1,k_2)\mathcal{M}_{n-1}(P,k_3,\cdots,k_n) + \cdots
\end{equation}
where $\mbox{Split}(t,k_1,k_2)$ is referred to as the splitting function, which is known, in gravity, to be
\begin{equation}
 \mbox{Split}(t,k_1,k_2) = f(t)\frac{[12]}{\langle 12\rangle}
\end{equation}
in terms of standard spinor helicity variables where $k_1\cdot k_2 = [12]\langle 12\rangle$ and $f(t)$ is a function of the collinear parameter $t$. In general, there are higher order terms in the collinear expansion but for the purposes of this work we do not consider those explicitly. It would be nice to understand these higher order collinear terms in this context, but we leave that to future exploration.

\paragraph{Boundary celestial algebras} Using both of these properties of bulk gravitational amplitudes, one can construct so-called ``celestial algebras" and ``celestial OPEs" in the boundary celestial CFT. Celestial amplitudes are constructed by Mellin transforming bulk amplitudes, thereby exchanging the energy label for a conformal dimension and resulting in an object that behaves like a conformal correlation function. We parametrize null momenta by 
\begin{equation}
q^\mu = \omega(1+z\bar{z},z+\bar{z},-i(z-\bar{z}),1-z\bar{z}) = \omega\hat{q}^\mu
\end{equation}
where $\omega$ is the energy and $(z,\bar{z})$ are complex coordinates on the celestial sphere. The polarization vectors that appear in the soft factors mentioned previously are given by $\varepsilon^+ = \partial_z\hat{q}$ and $\varepsilon^-=\partial_{\bar{z}}\hat{q}$ and the dot product of two vectors is given by $q_1\cdot q_2 = -2\omega_1\omega_2|z_{12}|^2$ which defines the relation to the spinor helicity variables used when discussing collinear limits. Note that the dot product of same helicity polarization vectors vanishes while that of opposite helicity polarization vectors is non-zero. This will be important as we consider collinear limits of same helicity operators in this work. We take the Mellin transform of a bulk amplitude as follows
\begin{equation}
\widetilde{\mathcal{M}}_n(z_1,\bar{z}_1,\Delta_1;\cdots;z_n,\bar{z}_n,\Delta_n) = \left[\prod_{i=1}^n\int_0^\infty d\omega_i \omega_i^{\Delta_i-1}\right]\mathcal{M}_n(k_1,\cdots,k_n)
\end{equation}
which gives us $\widetilde{\mathcal{M}}_n$, a celestial amplitude. Here the $\Delta_i$ are the conformal dimensions of the operators corresponding to each bulk field and $z_i,\bar{z}_i$ denote the location that they puncture the celestial sphere. 

The soft and collinear properties in the bulk have also been transformed to the boundary and are fairly well understood. Bulk soft theorems are transformed to ``conformally soft theorems" on the boundary which correspond to taking certain limits of the conformal dimension to generate the insertion of a ``soft operator". This is not surprising, since different values of $\Delta$ essentially correspond to different powers of $\omega$ which are the terms in a soft expansion. Letting $G_{\Delta}^{\pm}$ represent positive/negative helicity graviton operators in the CFT, one defines the conformally soft gravitons as~\cite{Guevara:2021abz}
\begin{equation}
H^k(z,\bar{z}) = \lim_{\varepsilon\rightarrow 0}\varepsilon G_{k+\varepsilon}^+(z,\bar{z}), \ \ k=2,1,0,-1.
\end{equation}
The $k=2$ operator is not physics, so $k=1$ is the one corresponding to the leading soft theorem, and $k=0$ is subleading and $k=-1$ is sub-subleading. It is relevant to point out, that we do not have to stop there and can clearly define operators for $k<-1$, and their interpretation is part of the motivation for this work. This full set of operators is referred to as the ``infinite tower" of soft operators in celestial CFTs.

If we instead look at how collinear limits are manifest on the boundary, we find that they correspond to OPEs.\cite{Fan:2019emx, Pate:2019lpp} This is also not surprising, since the OPE in a CFT requires bringing two operators to the same point on the sphere which is realized in the bulk as taking the corresponding momenta to be collinear. In practice, to take the collinear limit on the boundary one changes the integration variables in the corresponding Mellin integrals from $\omega_i,\omega_j$ to $\omega_P = \omega_i+\omega_j$ and $t=\frac{\omega_i}{\omega_P}$ and then takes $z_{ij}$ to be small which results in an expansion involving the aforementioned splitting function. Performing a Taylor expansion of the integrand with respect to $t$ results in the singular part of the OPE at tree level
\begin{equation}
G_{\Delta_1}^+(z_1,\bar{z}_1)G_{\Delta_2}^+(z_2,\bar{z}_2)\sim -\frac{\kappa}{2z_{12}}\sum_{n=0}^\infty B(\Delta_1-1+n,\Delta_2-1)\frac{\bar{z}_{12}^{n+1}}{n!}\bar{\partial}^n G_{\Delta_1+\Delta_2}^+(z_2,\bar{z}_2).
\end{equation}
One can then obtain the OPE of the soft gravitons by carefully taking the limit of this to obtain
\begin{equation}
H^k(z_1,\bar{z}_1)H^\ell(z_2,\bar{z}_2) \sim -\frac{\kappa}{2z_{12}}\sum_{n=0}^{1-k}\begin{pmatrix} 2-k-\ell-n\cr 1-\ell\end{pmatrix}\frac{\bar{z}_{12}^{n+1}}{n!}\bar{\partial}^n H^{k+\ell}(z_2,\bar{z}_2)
\end{equation}
from which one derives the soft algebra by considering the mode expansions. For the purposes of this paper, the OPE data will suffice. In particular, we will concentrate on the leading term in the OPE which means the aforementioned expansion in $z_{ij}$ does not matter for our purposes. 

The OPE and algebra appear to naturally give rise to an infinite tower of soft operators, where the more subleading operators are generated by the less subleading ones. In particular, we see the following: 
\begin{enumerate}
\item If we take two leading soft gravitons, their OPE is null. 
\item If we take a leading and subleading soft graviton, the leading term in the OPE is a leading soft graviton. Two subleading soft gravitons, gives a subleading soft graviton. So, we see that the first two form a closed set.
\item If we take a leading and sub-subleading soft graviton, we get a subleading soft graviton.
\item A subleading and sub-subleading soft graviton gives a sub-subleading soft graviton.
\item Two sub-subleading soft gravitons gives a sub$^{(3)}$-leading soft graviton. 	
\end{enumerate}
These statements are not new and were first introduced in~\cite{Guevara:2021abz}. The first point above is often said to be equivalent to the following statement
\begin{center}
\textbf{the leading soft theorem commutes}	
\end{center}
which is an important motivator for this work. The last point above shows that by incorporating the sub-subleading soft graviton, we are able to continue indefinitely and get the entire tower of sub$^{(n)}$-leading soft theorems. However, we know that beyond sub-subleading order, there is not a nice factorizable expression for the soft theorem.\footnote{Please contact the author if you would like to see what the sub$^{(3)}$-leading soft factor looks like for a generic gravitational amplitude. Since it is a long expression and does not have direct implications for this work, it has not been included.} A part of the soft expansion at these orders have been identified as ``quasi-soft theorems" but that is not the entire term in the expansion.\cite{Li:2018gnc} 

While the origin of these statements comes from the bulk scattering amplitudes, they have been solidified by using aspects of 2D CFT, namely using conformal symmetry. As far as we know, similar concrete statements about how to use lower order soft factors to generate higher order soft terms, have not been made at the bulk level. In this paper we aim to try and understand where these statements might arise in the context of bulk gravitational scattering amplitudes. In particular, we attempt to understand the following:
\begin{center}
\textbf{If the soft expansion fails, in general, to factorize beyond sub-subleading order, what do the $k<-1$ soft operators and associated conformally soft theorems correspond to in the bulk theory?}
\end{center}
In what follows, we try to answer this by performing the associated computations in a 4D theory of gravity and attempting to match the boundary results.

\section{Defining the double soft limit}\label{sec:doublesoft}
In order to make statements about how these boundary characteristics show up in the bulk, we define a double soft limit. In~\cite{Klose:2015xoa} they have a detailed discussion of how to compute consecutive and simultaneous double soft limits. It is natural to ask whether we can use their results to understand the boundary results from a bulk perspective. Therefore, we review this first. 

They have the following procedure to extract each order of the consecutive soft limit. First they define it as 
\begin{eqnarray}
& & \texttt{CSL}(1^{h_1},2^{h_2})\mathcal{M}_n(3,\cdots,n+2)  =  \lim_{\epsilon_1\rightarrow 0}\lim_{\epsilon_2\rightarrow 0}\mathcal{M}_{n+2}(\epsilon_1q_1^{h_1},\epsilon_2 q_2^{h_2},3,\cdots,n+2)\cr
& = & \left[\frac{1}{\epsilon_2}S^{(0)}(2^{h_2}) + S^{(1)}(2^{h_2}) + \epsilon_2S^{(2)}(2^{h_2}) +\cdots\right] \left[\frac{1}{\epsilon_1}S^{(0)}(1^{h_1}) + S^{(1)}(1^{h_1}) + \epsilon_1S^{(2)}(1^{h_1}) +\cdots\right]\cr
& \times & \mathcal{M}_n(3,\cdots,n+2)
\end{eqnarray}
where $h_1,h_2$ correspond to the helicities of the soft particles and we have chosen to work with an $n+2$ particle amplitude since we are taking two of them soft. Then, they set $\epsilon_1=\epsilon_2=\epsilon$ and organize the expansion in terms of the total power of $\epsilon$ which defines the leading term \texttt{CSL}$^{(0)}$, subleading term \texttt{CSL}$^{(1)}$ and so on. They further define symmetric and anti-symmetric combinations of this double soft limit as follows
\begin{eqnarray}
\texttt{sCSL}(1^{h_1},2^{h_2})\mathcal{M}_n(3,\cdots,n+2) & = & \frac{1}{2}\left\{\lim_{\epsilon_1\rightarrow 0},\lim_{\epsilon_2\rightarrow 0}\right\}\mathcal{M}_{n+2}(\epsilon_1q_1^{h_1},\epsilon_2q_2^{h_2},3,\cdots,n+2)\cr
\texttt{aCSL}(1^{h_1},2^{h_2})\mathcal{M}_n(3,\cdots,n+2) & = & \frac{1}{2}\left[\lim_{\epsilon_1\rightarrow 0},\lim_{\epsilon_2\rightarrow 0}\right]\mathcal{M}_{n+2}(\epsilon_1q_1^{h_1},\epsilon_2q_2^{h_2},3,\cdots,n+2).
\end{eqnarray}
na\"ively, one might suggest using their conventions exactly. It even seems like the anti-symmetric consecutive soft limit already has the interpretation as a commutator. However, in their paper the authors showed that the anti-symmetric combinations all vanished and the symmetric combinations did not look like the expected soft theorems, even if one takes the collinear limit. Therefore, to make contact with our boundary results, we need to consider some other object that is constructed from these double soft limits. 

Before we proceed, we will outline some notation that we will use in the following sections as we explore what the boundary description might be. We will write the double soft factor as 
\begin{equation}
S^{(m)}(1^+)S^{(n)}(2^+) \equiv \mathbb{S}_{12}^{(m,n)}.	
\end{equation}
Given the way the soft expansions stack, this order is understood to be the case where we take particle 1 soft first and then particle 2.\footnote{The way it is written may seem rather counterintuitive because it looks like $S^{(n)}(2^+)$ acts first, however when we perform the expansion rigorously, we see that this order appears when particle 1 is soft first.} If we were to write this out explicitly, we would see that $S^{(m)}(1^+)$ has dependence on $2^+$ which means that when it contains differential operators (i.e beyond leading order), it will act on $S^{(n)}(2^+)$ non-trivially. This is crucial if we want to understand the relationship to single soft limits. We then note that switching the particle order, i.e taking particle 2 soft first and then particle 1, is related by the following
\begin{equation}
\mathbb{S}_{12}^{\ast(m,n)}\equiv \mathbb{S}_{21}^{(m,n)} = 	\mathbb{S}_{12}^{(m,n)}(1\leftrightarrow 2).
\end{equation}
The polarization of the collinear particle in our discussion will be denoted $\varepsilon$. Each of these pairs, along with their associated collinear limits are calculated in appendix~\ref{app:doublesoft}. 

\section{Appearance of the double soft factors}\label{sec:usingdouble}
We will now explore some ways in which we can use these double soft limits to construct quantities that we can compare to what we expect from the boundary data.
\subsection{An easy answer: power matching}
We first note that the boundary results coming from the OPE can be heuristically produced by taking a collinear limit and then a soft limit and matching powers of $\epsilon$ to a simplistic double soft expansion. We will outline that logic first. 

Since we need to consider two soft particles, we will start with an $n+2$ particle amplitude and write the combination of soft expansions as follows
\begin{eqnarray}\label{eq:powermatching}
\lim_{\epsilon_1,\epsilon_2\rightarrow 0}\mathcal{M}_{n+2} & = & \left(\frac{1}{\epsilon_1}S_1^{(0)}+S_1^{(1)}+\epsilon_1S_1^{(2)}\right)\left(\frac{1}{\epsilon_2}S_2^{(0)}+S_2^{(1)}+\epsilon_2S_2^{(2)}\right)\mathcal{M}_n+\cdots\cr
& = & \frac{1}{\epsilon_1\epsilon_2}S_1^{(0)}S_2^{(0)}\mathcal{M}_n + \left[\frac{1}{\epsilon_1}S_1^{(0)}S_2^{(1)}+\frac{1}{\epsilon_2}S_1^{(1)}S_2^{(0)}\right]\mathcal{M}_n\cr
& + & \left[\frac{\epsilon_2}{\epsilon_1}S_1^{(0)}S_2^{(2)} + S_1^{(1)}S_2^{(1)} + \frac{\epsilon_1}{\epsilon_2}S_1^{(2)}S_2^{(0)}\right]\mathcal{M}_n\cr
& + & \left[\epsilon_1 S_1^{(2)}S_2^{(1)}+\epsilon_2 S_1^{(1)}S_2^{(2)}\right]\mathcal{M}_n + \epsilon_1\epsilon_2 S_1^{(2)}S_2^{(2)}\mathcal{M}_n+\cdots
\end{eqnarray}
where we have expanded and organized based on combined powers\footnote{There will be more contributions at order $\epsilon$ since we can multiply the order $\epsilon^2$ term with the leading soft term. However, since we do not have nice factorization of a generic amplitude at that order in the soft expansion, we do not include those contributions in this heuristic discussion.} of $\epsilon_1,\epsilon_2$. If we take the collinear limit of two particles first, and then take the soft limit of that collinear particle we will get a usual soft expansion which we can write as
\begin{equation}
\lim_{\epsilon\rightarrow 0}\lim_{1||2}\mathcal{M}_{n+2} = \mbox{Split}(t,q_1,q_2)\left(\frac{1}{\epsilon}S_P^{(0)}+S_P^{(1)}+\epsilon S_P^{(2)}\right)\mathcal{M}_n + \cdots.
\end{equation}
Now we take the limit of the first expression where $\epsilon_1=\epsilon_2=\epsilon$ and perform a rudimentary matching on the two sides of 
\begin{equation}
\lim_{\epsilon_1=\epsilon_2}\lim_{\epsilon_1,\epsilon_2\rightarrow 0}\mathcal{M}_{n+2} = \lim_{\epsilon\rightarrow 0}\lim_{1||2}\mathcal{M}_{n+2}.
\end{equation}
Since there is no $\epsilon^{-2}$ term on the right hand side, the conclusion is that the OPE of $S_1^{(0)}, S_2^{(0)}$ is null. Likewise, matching the other powers on each side we can see that leading with subleading will give leading, leading with sub-subleading and subleading with subleading will give subleading and subleading with sub-subleading gives sub-subleading. These relations are all consistent with the OPE statements on the boundary.

This is not surprising, because this is suggestive of taking the residues at different powers of the energy if we remember that we can equivalently write this as an expansion in the energies $\omega_i$ rather than the small parameters $\epsilon_i$.\cite{Guevara:2019ypd} However, if we look at the left hand side we  realize that while they are related to the OPE statements, they are not exactly the same. For example in the $\mathcal{O}(1)$ term on the left hand side, we have a combination of terms pairing leading with sub-subleading as well as subleading with subleading. This is not what appears in the OPE. Part of this is a result of performing this computation on the boundary, where orders in soft expansions are represented by certain integer values of the conformal weights of the associated operators, we are able to simply choose the soft operators that appear by fixing the weights. In this case, a similar construction in the bulk has more ambiguities because we do not have that power to choose. Additionally, as we will see later, the consecutive double soft expansion does not organize in this way since there are additional factors of $\epsilon$ that need to be taken into consideration. In what follows, we will see if there are other methods by which we can extract the information we need by defining some kind of convention for this limiting process.

\subsection{A schematic viewpoint}\label{sec:scheme}
Before we get to the actual computations of these explicit double soft limits, we should point out two things that will be relevant to extracting the terms we care about. Let us write the soft factors as 
\begin{equation}
S_\ell^{(i)} = \sum_{a=1,a\neq\ell}^{n+2}\frac{S_{\ell,a}^{(i)}}{q_\ell\cdot k_a}
\end{equation}
where we are assuming that we are starting with an $n+2$ particle amplitude, $\ell$ denotes the particle that is taken soft and $S_{\ell,a}^{(i)}$ is generically a differential operator where the derivatives are with respect to $k_a,\varepsilon_a$. We can then see that the double soft factors that we obtain by taking consecutive soft limits look like
\begin{equation}
S_\ell^{(i)}S_{\ell'}^{(j)}\mathcal{M}_n = \left(\sum_{a=1,a\neq\ell}^{n+2}\frac{S_{\ell,a}^{(i)}}{q_\ell\cdot k_a}\right)\left(\sum_{b=1,b\neq\ell,\ell'}^{n+2}\frac{S_{\ell',b}^{(j)}}{q_{\ell'}\cdot k_b}\right)\mathcal{M}_n
\end{equation}
where we see that since we took particle $\ell$ to be soft first, the first sum contains $\ell'$. Since we are primarily concerned with matching to leading OPE data on the boundary, the most important piece of this will be the leading collinear singularity when $\ell || \ell'$. Looking at the above, we can see that comes from the term when $a=\ell'$. Therefore, to extract the leading collinear data we only need
\begin{equation}
S_\ell^{(i)}S_{\ell'}^{(j)}\mathcal{M}_n = \left(\frac{S_{\ell,\ell'}^{(i)}}{q_\ell\cdot q_{\ell'}}\right)\left[\sum_{b=1,b\neq\ell,\ell'}^{n+2}\frac{S_{\ell',b}^{(j)}}{q_{\ell'}\cdot k_b}\right]\mathcal{M}_n + \cdots 
\end{equation}
Furthermore, we can also see that since $\mathcal{M}_n$ has no dependence on $q_\ell,q_{\ell'}$, the differential operator $S_{\ell,\ell'}$ will only act on the other soft factor, thereby giving a nice procedure to extract this leading term. 

We can also infer some schematic properties of the derivative structure of the result. Since $S_{\ell,\ell'}^{(i)}$ does not act on $\mathcal{M}_n$, we see that the result of this action will have the same derivative structure as $S_{\ell',b}^{(j)}$. It cannot have fewer derivatives since $S_{\ell,\ell'}^{(i)}$ will not act on the derivative part of $S_{\ell',b}^{(j)}$ so it will have exactly the same derivative structure. This tells us that for the singular part of this expression, if, for instance, we let $i=0$ and $j=2$, the result will have the structure of the sub-subleading soft factor when we actually expect the leading soft factor based on the boundary OPEs. This is already compelling evidence that just the combination of soft factors will not be enough but rather that we need a commutator or some other related structure instead. 

A na\"ive commutator to consider is the following 
\begin{equation}
\left[S_\ell^{(i)}, S_{\ell'}^{(j)}\right] \equiv \left[\left(\sum_{a=1,a\neq\ell}^{n+2}\frac{S_{\ell,a}^{(i)}}{q_\ell\cdot k_a}\right)\left(\sum_{b=1,b\neq\ell,\ell'}^{n+2}\frac{S_{\ell',b}^{(j)}}{q_{\ell'}\cdot k_b}\right),\left(\sum_{b=1,b\neq\ell,\ell'}^{n+2}\frac{S_{\ell',b}^{(j)}}{q_{\ell'}\cdot k_b}\right)\left(\sum_{a=1,a\neq\ell}^{n+2}\frac{S_{\ell,a}^{(i)}}{q_\ell\cdot k_a}\right)\right].
\end{equation}
The first term in this commutator is what we computed above. The second term is one that does not make much physical sense since it would imply that we first took $\ell'$ soft, ignoring particle $\ell$ as a hard contribution, and then took particle $\ell$ soft, including $\ell'$ as a hard contribution when it was already soft. The singular part of that term, will be when $a=\ell'$ but in that case, $S^{(i)}_{\ell,\ell'}$ will vanish when acting on $\mathcal{M}_n$ and therefore this commutator is just exactly equal to the first term we expanded above. We can, therefore, conclude that we need to define a different notion of the commutator if we want to match the expected results. 

In what follows, we will show this schematic argument explicitly and use that to compute candidates for what the bulk realization of the boundary algebra could be.

\section{Explicit double soft factors}\label{sec:explicitsoft}
The first, and easiest, option is to consider the double soft limit and take the collinear limit of it. In general, one does not expect that soft and collinear limits commute so we expect that this should be different from taking the collinear limit first and then the soft limit of the resulting particle. 

There are nine possible options here which we list below along with some relevant discussion about each. We have written these conveniently in terms of the known soft factors and their derivatives to help make identifications easy. The general derivatives of soft factors are computed in appendix~\ref{app:derivatives} and the double soft factors are computed explicitly in appendix~\ref{app:doublesoft}. We first look at the double leading soft theorem
\begin{equation}
\lim_{1||2}\mathbb{S}_{12}^{(0,0)}  =  \frac{1}{\epsilon^2}\frac{1}{t(1-t)}S_P^{(0)}S_P^{(0)} -\frac{1}{\epsilon}\frac{2}{t}\frac{\bar{z}_{12}}{z_{12}} S_P^{(0)}.
\end{equation}
and we see that it does not vanish. That should not be surprising, seeing as the leading soft factor is a scalar so there is no reason for a product of scalars to vanish. We could look at this two ways. We see that, due to the additional expansion of the soft factor itself under the second soft limit, the leading term in the $\frac{1}{\epsilon}$ expansion is not a single particle operator but rather the product of two leading soft factors. Therefore, we could choose a prescription where we look at the leading single particle term in the $\frac{1}{\epsilon}$ expansion and in that case, we could conclude here that it vanishes. Alternatively, we could say that we want to look at the leading collinear term, which is a single particle term since it is just a leading soft factor and that would be one motivation to consider some combination like a commutator. 

If we take one leading and one subleading, we have two different orderings
\begin{eqnarray}
\lim_{1||2}\mathbb{S}_{12}^{(0,1)} & = & \frac{1}{\epsilon}\frac{1}{t}S_P^{(0)}S_P^{(1)}-\frac{2(1-t)}{t}\frac{\bar{z}_{12}}{z_{12}}  S_P^{(1)}\cr
\lim_{1||2}\mathbb{S}_{12}^{(1,0)} & = & \frac{1}{\epsilon}\frac{2}{(1-t)}\frac{\bar{z}_{12}}{z_{12}}S_P^{(0)}  + \frac{1}{\epsilon}\frac{1}{(1-t)}S_P^{(0)}S_P^{(1)} - \frac{1}{\epsilon}\frac{1}{1-t}\omega_P^2|z_{12}|^2 \partial_{z_P}^2S_P^{(0)}.
\end{eqnarray}
We first see that the two orderings are different which is not surprising given that the leading and subleading soft factors have different numbers of derivatives and generically differential operators do not commute. We see that the leading collinear limit gives us the subleading soft factor for the first ordering and the leading soft factor in the second. As per the OPE discussion, we expect to get the leading soft theorem which means if we consider this set of limits, only one ordering gives us the answer we expect even though there is no reason to implicitly choose an order. Additionally, from our discussion in the previous section, the fact that we get leading in one case and subleading in the other should not be surprising since the result takes the derivative structure of the second soft factor. It is possible that we could get around the ordering issue here if we chose to look at the leading order in $\frac{1}{\epsilon}$ because in that case, the first ordering has no single particle contribution while the second case does. 

For the leading and sub-subleading we also have two different orderings
\begin{eqnarray}
\lim_{1||2}\mathbb{S}_{12}^{(0,2)} & = & \frac{1-t}{t}S_P^{(0)}S_P^{(2)}-\epsilon\frac{2(1-t)^2}{t}\frac{\bar{z}_{12}}{z_{12}} S_P^{(2)}\cr 
\lim_{1||2}\mathbb{S}_{12}^{(2,0)} & = &  -\frac{1}{\epsilon}\frac{t}{(1-t)^2}\omega_P^2|z_{12}|^2\partial_{z_P}^2S_P^{(0)} +\frac{2t}{1-t} \omega_P\bar{z}_{12}\partial_{z_P}S_P^{(1)}  + \frac{t}{1-t}S_P^{(0)}S_P^{(2)}\cr
& - & \frac{t}{(1-t)}\omega_P^2|z_{12}|^2\partial_{z_P}^2S_P^{(1)} + \frac{4t}{(1-t)}\omega_P^4|z_{12}|^4\sum_{a=3}^{n+2}\frac{(\varepsilon_P\cdot k_a)^4}{(P\cdot k_a)^4}
\end{eqnarray}
Once again, we see that there are different answers for each ordering. In this case, we expect to get the subleading soft factor but that only appears in the second line with a derivative and not at leading order in the collinear expansion. The fact that we get the sub-subleading soft factor at leading order in the collinear expansion is not surprising, given our argument about ordering of derivatives in the previous section. Since here we do not get the expected answer at all, it seems that even picking a prescription where we filter by $\epsilon$ expansion will not help.

Next we look at when both are subleading which has only one ordering
\begin{eqnarray} 
\lim_{1||2}\mathbb{S}_{12}^{(1,1)} & = & \frac{4\bar{z}_{12}}{z_{12}} S_P^{(1)} - 2\omega_P\bar{z}_{12}  \partial_{z_P}S_P^{(1)} - \omega_P^2|z_{12}|^2\partial_{z_P}^2S_P^{(1)} \cr
& - & \sum_{a,b=3}^{n+2}\frac{(\varepsilon_P\cdot k_a) }{P\cdot k_a} \frac{(\varepsilon_P\cdot k_b)}{P\cdot k_b} \left(\varepsilon_{P,\nu} P_{\rho}\right)\left(\varepsilon_{P,\beta}P_{\lambda}\right) J_a^{\rho\nu}J_b^{\lambda\beta}
\end{eqnarray}
Here at leading order in the collinear limit we get the expected result of the subleading soft factor. Once again, this is not surprising. If we kept the $\epsilon$ expansion prescription suggested in the first two cases, it would work here as well.

Next we look at one subleading and one sub-subleading which has two orderings
\begin{eqnarray}
\lim_{1||2}\mathbb{S}_{12}^{(1,2)} & = & 6\epsilon(1-t)\frac{\bar{z}_{12}}{z_{12}} S_P^{(2)} - 4\epsilon(1-t)\omega_P\bar{z}_{12} \partial_{z_P}S_P^{(2)} - \epsilon(1-t)\omega_P^2|z_{12}|^2 \partial_{z_P}^2S_P^{(2)}\cr 
& + & \frac{i\epsilon}{2}(1-t) \sum_{a,b=3}^{n+2}\frac{(\varepsilon_P\cdot k_a) }{P\cdot k_a}  \frac{\left(\varepsilon_{P,\nu}P_{\rho}J_a^{\rho\nu}\right)\left(\varepsilon_{P,\alpha}P_{\lambda}J_b^{\lambda\alpha}\right)\left(P_{\tau}\varepsilon_{P,\beta}J_b^{\tau\beta}\right)}{P\cdot k_b}\cr
\lim_{1||2}\mathbb{S}_{12}^{(2,1)} & = & -\frac{2t}{1-t}\frac{\bar{z}_{12}}{z_{12}}S_P^{(1)} + \frac{2t}{1-t}\omega_P\bar{z}_{12}\partial_{z_P}S_P^{(1)}  - \frac{t}{1-t}\omega_P^2|z_{12}|^2\partial_{z_P}^2S_P^{(1)}\cr
& - & 2\epsilon t\omega_P^2|z_{12}|^2\partial_{z_P}S_P^{(2)} - 2\epsilon t\omega_P^3|z_{12}|^2\bar{z}_{12}\partial_{z_P}^2S_P^{(1)} \cr
& + &  \frac{i\epsilon t}{2}\sum_{a,b=3}^{n+2}\frac{(\varepsilon_P\cdot k_b)}{P\cdot k_a} \frac{ \left(\varepsilon_{P,\mu}P_{\rho}J_a^{\rho\mu}\right)\left(\varepsilon_{P,\nu}P_{\sigma}J_a^{\sigma\nu}\right)\left(\varepsilon_{P,\beta}P_{\lambda}J_b^{\lambda\beta}\right)}{P\cdot k_b} \cr
& - & 4i\epsilon t\omega_P^4|z_{12}|^4\sum_{a=3}^{n+2} \frac{(\varepsilon_P\cdot k_a)^3 \left(\varepsilon_{P,\beta}P_{\lambda}J_a^{\lambda\beta}\right)}{(P\cdot k_a)^4} 
\end{eqnarray}
In this case the expected answer is the sub-subleading soft factor which appears in the first ordering at leading order in the collinear limit but not in the second ordering. Once again, we see that ordering matters but we have one case where the answer comes out as expected. This can be picked out using the $\epsilon$ expansion prescription. 

Lastly we look at the case when both are sub-subleading
\begin{eqnarray}
\lim_{1||2}\mathbb{S}_{12}^{(2,2)} & = & -6\epsilon t\frac{\bar{z}_{12}}{z_{12}}S_P^{(2)} + 4\epsilon t\omega_P\bar{z}_{12}\partial_{z_P}S_P^{(2)} -  \epsilon t\omega_P^2|z_{12}|^2\partial_{z_P}^2S_P^{(2)}\cr
& - & 2\epsilon^2t(1-t)\omega_P^2\bar{z}_{12}^2 \sum_{a=3}^{n+2}\frac{1}{(P\cdot k_a)^2}\left(P_{\lambda}\varepsilon_{P,\alpha}J_a^{\lambda\alpha}\right)  \left(P_{\tau}\varepsilon_{P,\beta}J_a^{\tau\beta}\right)\cr 
& + & 4\epsilon^2t(1-t)\omega_P^3|z_{12}|^2\bar{z}_{12} \sum_{a=3}^{n+2}\frac{(\varepsilon_P\cdot k_a)}{(P\cdot k_a)^3}\left(P_{\lambda}\varepsilon_{P,\alpha}J_a^{\lambda\alpha}\right)  \left(P_{\tau}\varepsilon_{P,\beta}J_a^{\tau\beta}\right)\cr
& - & 2\epsilon^2t(1-t)\omega_P^4|z_{12}|^4\sum_{a=3}^{n+2}\frac{(\varepsilon_P\cdot k_a)^2}{(P\cdot k_a)^4}\left(P_{\lambda}\varepsilon_{P,\alpha}J_a^{\lambda\alpha}\right)  \left(P_{\tau}\varepsilon_{P,\beta}J_a^{\tau\beta}\right)\cr 
& + & \frac{\epsilon^2 t(1-t)}{4} \sum_{a,b=3}^{n+2}\frac{1}{P\cdot k_a} \frac{\left(P_{\rho}\varepsilon_{P,\mu}J_a^{\rho\mu}\right)\left(P_{\sigma}\varepsilon_{P,\nu}J_a^{\sigma\nu}\right)\left(P_{\lambda}\varepsilon_{P,\alpha}J_b^{\lambda\alpha}\right)\left(P_{\tau}\varepsilon_{P,\beta}J_b^{\tau\beta}\right)}{P\cdot k_b} \cr
& + & i\epsilon^2 t(1-t)\omega_P\bar{z}_{12}\sum_{a=3}^{n+2} \frac{ \left(P_{\sigma}\varepsilon_{P,\nu}J_a^{\sigma\nu}\right)\left(P_{\lambda}\varepsilon_{P,\alpha}J_a^{\lambda\alpha}\right)\left(P_{\tau}\varepsilon_{P,\beta}J_a^{\tau\beta}\right)}{(P\cdot k_a)^2} \cr
& - & i\epsilon^2t(1-t)\omega_P^2|z_{12}|^2\sum_{a=3}^{n+2} \frac{(\varepsilon_P\cdot k_a)\left(P_{\sigma}\varepsilon_{P,\nu}J_a^{\sigma\nu}\right)\left(P_{\lambda}\varepsilon_{P,\alpha}J_a^{\lambda\alpha}\right)\left(P_{\tau}\varepsilon_{P,\beta}J_a^{\tau\beta}\right)}{(P\cdot k_a)^3}
\end{eqnarray}
and we see that at leading order in the collinear limit we have the sub-subleading soft factor which is not what we expected from the boundary perspective but is the na\"ive expectation given the derivative structure. We expected to get something that was higher order in derivatives since it should give something related to the sub$^{(3)}$-leading soft theorem.

\textbf{If we restrict ourselves to just the leading and subleading soft factors, then picking the leading $\frac{1}{\epsilon}$ term along with the leading collinear singularity, gives us consistency with the boundary statements.} However, once we include the sub-subleading soft factor this breaks down and we do not get consistency anymore. The statements from~\cite{Guevara:2021abz, Himwich:2023njb} suggest that we should be able to consistently generate all the infinite terms somehow so we are still left with the puzzle of how to incorporate the sub-subleading terms. 

\subsection{$\epsilon$ expansion}
From the above, we notice that each of these double soft factors has contributions at various orders in $\epsilon$. Therefore, we can also take the sum of all of them and then organize that with respect to powers in $\epsilon$. This will look like 
\begin{eqnarray}
\mathtt{CSL}_{12}\mathcal{M}_n & = & \frac{1}{\epsilon^2} \mathbb{S}_{12}^{(0,0),-2}\mathcal{M}_{n} \cr
& + & \frac{1}{\epsilon}\left(\mathbb{S}_{12}^{(0,0),-1}  +\mathbb{S}_{12}^{(0,1),-1} + \mathbb{S}_{12}^{(1,0),-1} + \mathbb{S}_{12}^{(2,0),-1}\right)\mathcal{M}_{n} \cr
& + & \left(\mathbb{S}_{12}^{(0,1),0}  + \mathbb{S}_{12}^{(0,2),0} + \mathbb{S}_{12}^{(1,1),0} +  \mathbb{S}_{12}^{(2,0),0}  + \mathbb{S}_{12}^{(2,1),0}\right)\mathcal{M}_{n} \cr
& + & \epsilon\left(\mathbb{S}_{12}^{(0,2),1}+\mathbb{S}_{12}^{(1,2)}+\mathbb{S}_{12}^{(2,1),1}+\mathbb{S}_{12}^{(2,2),1}\right)\mathcal{M}_{n} \cr
& + & \epsilon^2\mathbb{S}_{12}^{(2,2),2}\mathcal{M}_{n} + \cdots
\end{eqnarray}
where the $\cdots$ refer to additional terms that will come from higher order terms in each of the individual soft expansions, i.e beyond the sub-subleading order and we have denoted $\mathbb{S}_{12}^{(i,j),\alpha}$ to be the coefficient of $\epsilon^{\alpha}$ in $\mathbb{S}_{12}^{(i,j)}$. If we look at the leading collinear limit of this we have 
\begin{equation}\label{eq:CSL}
\lim_{1||2}\mathtt{CSL}_{12} =  \frac{1}{\epsilon}\left(\frac{4t-2}{t(1-t)}\right)\frac{\bar{z}_{12}}{z_{12}}S_P^{(0)} - \frac{2(2t-1)^2}{t(1-t)}\frac{\bar{z}_{12}}{z_{12}}S_P^{(1)}-\epsilon \left(\frac{14t^2-10t+2}{t}\right)\frac{\bar{z}_{12}}{z_{12}}S_P^{(2)} +\cdots 
\end{equation}
There are a few things to notice here. 
\begin{enumerate}
\item We see that this does organize itself as a single soft expansion in the collinear momentum $P$. However, it is not as simple as the splitting function multiplying a soft expansion since the collinear parameter $t$ appears here rather non-trivially.
\item The terms at each order in $\epsilon$ are noticeably different from the na\"ive expansion in equation~\eqref{eq:powermatching} which comes from additional powers of the soft expansion parameter when performing the double soft expansion carefully. Therefore, we cannot immediately come up with the OPE relations that we are striving for because there is no particular prescription to choose a specific pair of soft factors and relate it to the collinear soft expansion.
\item The expectation from the boundary description is that if we are equipped with the leading, subleading and sub-subleading soft factors, we should be able to generate the entire tower i.e all the sub$^{(n)}$-leading soft factors. However, we see here that this collinear limit does not give any terms beyond order $\epsilon$ since there are no leading collinear terms in $\epsilon^2\mathbb{S}_{12}^{(2,2),2}$. Therefore, it looks like either we are unable to generate anything past sub-subleading order in this way or that we would need to include terms in the individual soft expansions that are beyond sub-subleading order to generate these. Neither of these options are consistent with our expectations.
\end{enumerate}
It should also be noted that we could have chosen the opposite ordering, $\mathtt{CSL}_{21}$. In that case, specifically for these leading collinear terms, we would just need to take $t\rightarrow 1-t$ to get the corresponding expansion. In the following section we explore whether any of these results can be framed in terms of commutators of soft limits.

\section{Commutators and anti-commutators}\label{sec:commutators}
Since the double soft factors on their own did not give us results that were fully consistent beyond the subleading level, we can look at a commutator. In subsection~\ref{sec:scheme} we saw that the na\"ive commutator had an ordering that did not make physical sense and thereby vanished when acting on the amplitude resulting in that commutator being equal to the double soft factors explored in the previous section. Here we opt for the following definition of the commutator
\begin{equation}
\mathtt{commSL}^{(m,n)}(1^+,2^+) = \lim_{1||2}\left(\mathbb{S}_{12}^{(m,n)}-\mathbb{S}_{21}^{(n,m)}\right) = \lim_{1||2}\left(\mathbb{S}_{12}^{(m,n)}-\mathbb{S}_{12}^{\ast(n,m)}\right)
\end{equation}
which allows for both terms to have physical meaning since they are genuinely two separate orderings of the soft expansion. First we can write it out in orders of $\epsilon$ so that we can easily extract the associated terms order by order. Using the results above we have 
\begin{equation}
\mathtt{commSL}^{(0,0)}(1^+,2^+) =\frac{1}{\epsilon}\left[\frac{2}{(1-t)}-\frac{2}{t}\right]\frac{\bar{z}_{12}}{z_{12}} S_P^{(0)}.
\end{equation}
At order $\epsilon^{-2}$ this vanishes as expected. Only when we look at the next order in the double soft limit, we have a non-vanishing term. Next we have 
\begin{equation}
\mathtt{commSL}^{(0,1)}(1^+,2^+) =  -\frac{1}{\epsilon}\frac{2}{t}\frac{\bar{z}_{12}}{z_{12}}S_P^{(0)}   + \frac{1}{\epsilon}\frac{1}{t}\omega_P^2|z_{12}|^2 \partial_{z_P}^2S_P^{(0)}-\frac{2(1-t)}{t}\frac{\bar{z}_{12}}{z_{12}}  S_P^{(1)}
\end{equation}
At order $\epsilon^{-1}$, the leading term in the collinear limit is the leading soft factor which is what we expect. 
\begin{eqnarray}
\mathtt{commSL}^{(0,2)}(1^+,2^+) & = & \frac{1}{\epsilon}\frac{(1-t)}{t^2}\omega_P^2|z_{12}|^2\partial_{z_P}^2S_P^{(0)} +\frac{2(1-t)}{t} \omega_P\bar{z}_{12}\partial_{z_P}S_P^{(1)} +\frac{1-t}{t}\omega_P^2|z_{12}|^2\partial_{z_P}^2S_P^{(1)} \cr
& - & \frac{4(1-t)}{t}\omega_P^4|z_{12}|^4\sum_{a=3}^{n+2}\frac{(\varepsilon_P\cdot k_a)^4}{(P\cdot k_a)^4} -\epsilon\frac{2(1-t)^2}{t}\frac{\bar{z}_{12}}{z_{12}} S_P^{(2)}
\end{eqnarray}
Unfortunately, this pattern starts to break down. The leading order in $\epsilon$ is $\epsilon^{-1}$ which is two derivatives of the leading soft factor whereas here, we expect to get the subleading soft factor. This breaks down further if we look at the commutator of two subleading soft factors
\begin{eqnarray}
\mathtt{commSL}^{(1,1)}(1^+,2^+) & = &  - 4\omega_P\bar{z}_{12}  \partial_{z_P}S_P^{(1)} . 
\end{eqnarray}
We see that this is not zero but it is a derivative of the subleading soft factor whereas we expected to get the subleading soft factor. We have two more commutators left to compute 
\begin{eqnarray}
\mathtt{commSL}^{(1,2)}(1^+,2^+) & = & \frac{2(1-t)}{t}\frac{\bar{z}_{12}}{z_{12}}S_P^{(1)} + \frac{2(1-t)}{t}\omega_P\bar{z}_{12}\partial_{z_P}S_P^{(1)}  + \frac{1-t}{t}\omega_P^2|z_{12}|^2\partial_{z_P}^2S_P^{(1)}\cr
& + & 6\epsilon(1-t)\frac{\bar{z}_{12}}{z_{12}} S_P^{(2)} - 4\epsilon(1-t)\omega_P\bar{z}_{12} \partial_{z_P}S_P^{(2)} - \epsilon(1-t)\omega_P^2|z_{12}|^2 \partial_{z_P}^2S_P^{(2)}\cr 
& + & 2\epsilon (1-t)\omega_P^2|z_{12}|^2\partial_{z_P}S_P^{(2)} - 2\epsilon (1-t)\omega_P^3|z_{12}|^2\bar{z}_{12}\partial_{z_P}^2S_P^{(1)} \cr
& + & \frac{i\epsilon}{2}(1-t) \sum_{a,b=3}^{n+2}\frac{(\varepsilon_P\cdot k_a) }{P\cdot k_a}  \frac{\left(\varepsilon_{P,\nu}P_{\rho}J_a^{\rho\nu}\right)\left(\varepsilon_{P,\alpha}P_{\lambda}J_b^{\lambda\alpha}\right)\left(P_{\tau}\varepsilon_{P,\beta}J_b^{\tau\beta}\right)}{P\cdot k_b}\cr
& - &  \frac{i\epsilon (1-t)}{2}\sum_{a,b=3}^{n+2}\frac{(\varepsilon_P\cdot k_b)}{P\cdot k_a} \frac{ \left(\varepsilon_{P,\mu}P_{\rho}J_a^{\rho\mu}\right)\left(\varepsilon_{P,\nu}P_{\sigma}J_a^{\sigma\nu}\right)\left(\varepsilon_{P,\beta}P_{\lambda}J_b^{\lambda\beta}\right)}{P\cdot k_b} \cr
& + & 4i\epsilon (1-t)\omega_P^4|z_{12}|^4\sum_{a=3}^{n+2} \frac{(\varepsilon_P\cdot k_a)^3 \left(\varepsilon_{P,\beta}P_{\lambda}J_a^{\lambda\beta}\right)}{(P\cdot k_a)^4} .
\end{eqnarray}
At leading order in $\epsilon$ and the soft expansion, we get the subleading soft factor which is also not what we expected. Finally
\begin{eqnarray}
\mathtt{commSL}^{(2,2)}(1^+,2^+) & = & 6\epsilon (1-2t)\frac{\bar{z}_{12}}{z_{12}}S_P^{(2)} + 4\epsilon \omega_P\bar{z}_{12}\partial_{z_P}S_P^{(2)} + \epsilon (1-2t)\omega_P^2|z_{12}|^2\partial_{z_P}^2S_P^{(2)}\cr
& + & 8\epsilon^2t(1-t)\omega_P^3|z_{12}|^2\bar{z}_{12} \sum_{a=3}^{n+2}\frac{(\varepsilon_P\cdot k_a)}{(P\cdot k_a)^3}\left(P_{\lambda}\varepsilon_{P,\alpha}J_a^{\lambda\alpha}\right)  \left(P_{\tau}\varepsilon_{P,\beta}J_a^{\tau\beta}\right)\cr
& + & 2i\epsilon^2 t(1-t)\omega_P\bar{z}_{12}\sum_{a=3}^{n+2} \frac{ \left(P_{\sigma}\varepsilon_{P,\nu}J_a^{\sigma\nu}\right)\left(P_{\lambda}\varepsilon_{P,\alpha}J_a^{\lambda\alpha}\right)\left(P_{\tau}\varepsilon_{P,\beta}J_a^{\tau\beta}\right)}{(P\cdot k_a)^2} .
\end{eqnarray}
If we look at this at face value, it looks like the leading collinear limit gives us the sub-subleading soft factor which is not consistent with the boundary expectation. This is where we would have expected to get the universal part of the sub$^{(3)}$-leading soft term.

Once again, it looks like we have no consistent approach to defining a commutator that is consistent with the boundary expectations so we need to consider some other object. In particular, we see that this commutator does not even give the closed subalgebra of just the leading and subleading terms. This is surprising because a commutator as defined here, seems to na\"ively be consistent with the statement that the leading soft theorem commutes as well as with the notion of boundary algebras. However, what we find is that the OPE is not consistent with the combination of limits and the algebra is not consistent with the commutator so there must be something else missing when finding a bulk interpretation of the boundary soft algebras.

\subsection{Full commutator or anti-commutator order by order}
We could consider the full commutator (or anti-commutator), abandoning the notion of specific pairs of operators that is suggested by the boundary description, and analyze it order by order in the $\epsilon$ expansion. This would be more akin to the anti-symmetric consecutive soft limit that was considered in~\cite{Klose:2015xoa} but it is plausible that combined with the collinear limit, we see some different behavior. We can utilize the result in equation~\eqref{eq:CSL} and either add or subtract the opposite ordering. Focusing on just the leading collinear singularities we get 
\begin{eqnarray}
{}^{[ \ , \ ]}\mathtt{CSL}_{12}\mathcal{M}_n & = & \frac{2}{\epsilon}\left(\frac{4t-2}{t(1-t)}\right)\frac{\bar{z}_{12}}{z_{12}}S_P^{(0)}\mathcal{M}_n+ \epsilon \left(\frac{2(2t-1)(7t^2-7t+1)}{(1-t)t}\right)\frac{\bar{z}_{12}}{z_{12}}S_P^{(2)}\mathcal{M}_n +\cdots \cr
{}^{\{ \ , \ \}}\mathtt{CSL}_{12}\mathcal{M}_n & = &- \frac{4(2t-1)^2}{t(1-t)}\frac{\bar{z}_{12}}{z_{12}}S_P^{(1)}\mathcal{M}_n - \epsilon \left(\frac{2(3t^2-3t+1)}{(1-t)t}\right)\frac{\bar{z}_{12}}{z_{12}}S_P^{(2)}\mathcal{M}_n+\cdots
\end{eqnarray}
for the commutator and anti-commutator respectively. At first glance this actually seems promising, but in either case we have cancelled out one of the first two soft factors which is not what we want. We can thereby conclude that the consecutive soft limit in equation~\eqref{eq:CSL} would be better than either its commutator or anti-commutator but in all cases, we lose the information about which two soft factors were combined to give a collinear soft factor. 

\section{Gauge theory}\label{sec:gauge}
Given what we have seen in gravity, it is natural to ask whether the gauge theory analog looks different. The soft expansion for gauge theory is universal to second order and is given by~\cite{Weinberg:1965nx,Broedel:2014fsa}
\begin{eqnarray}
    \lim_{\epsilon_1\rightarrow 0}\mathcal{A}_{n+2} & = & \left[\frac{1}{\epsilon_1}\sum_{a=2}^{n+2}\frac{\varepsilon_1\cdot k_a}{q_1\cdot k_a}+\left(\frac{\varepsilon_{1,\mu}q_{1,\nu}J_2^{\mu\nu}}{q_1\cdot k_2}-\frac{\varepsilon_{1,\mu}q_{1,\nu}J_{n+2}^{\mu\nu}}{q_1\cdot k_{n+2}}\right)\right]\mathcal{A}_{n+1}\cr
    & = & \left(\frac{1}{\epsilon_1}S_1^{(0)}+S_1^{(1)}\right)\mathcal{A}_{n+1}
\end{eqnarray}
where we have assumed that the soft operator is positive helicity, as in the gravitational case. As in the gravity example, we can look at the leading collinear terms in one ordering of the double soft limit (since we know the other ordering is related via a replacement of labels)
\begin{eqnarray}
    \lim_{\epsilon_1,\epsilon_2\rightarrow 0}\mathcal{A}_{n+2} & = & \left[\frac{1}{\epsilon_1}\frac{\varepsilon_1\cdot q_2}{q_1\cdot q_2} +\frac{1}{\epsilon_2}\frac{\varepsilon_{1,\mu}q_{1,\nu}J_2^{\mu\nu}}{q_1\cdot q_2}\right]\left[\frac{1}{\epsilon_2}\sum_{b=3}^{n+2}\frac{\varepsilon_2\cdot k_b}{q_2\cdot k_b}+\left(\frac{\varepsilon_{2,\rho}q_{2,\sigma}J_3^{\rho\sigma}}{q_2\cdot k_3}-\frac{\varepsilon_{2,\rho}q_{2,\sigma}J_{n+2}^{\rho\sigma}}{q_2\cdot k_{n+2}}\right)\right]\mathcal{A}_{n} \cr
    & = & \left[\frac{1}{\epsilon_1\epsilon_2}\mathbb{S}_{12}^{(0,0)} + \frac{1}{\epsilon_1}\mathbb{S}_{12}^{(0,1)} + \frac{1}{\epsilon_2^2}\mathbb{S}_{12}^{(1,0)} + \frac{1}{\epsilon_2}\mathbb{S}_{12}^{(1,1)}\right]\mathcal{A}_n+\cdots 
\end{eqnarray}
We have used the same notation here for the double soft factors as was used previously. We can write each of these terms in the collinear limit using the same notation as before and we find
\begin{eqnarray}
\lim_{1||2}\mathbb{S}_{12}^{(0,0)} & = & \frac{1}{t(1-t)\omega_Pz_{12}}\sum_{b=3}^{n+2}\frac{\varepsilon\cdot k_b}{P\cdot k_b} = \frac{1}{t(1-t)\omega_Pz_{12}}S_P^{(0)}\cr
\lim_{1||2}\mathbb{S}_{12}^{(0,1)} & = & \frac{1}{t\omega_Pz_{12}}\left(\frac{\varepsilon_{\mu}P_{\nu}J_3^{\mu\nu}}{P\cdot k_3}-\frac{\varepsilon_{2}P_{\nu}J_{n+2}^{\mu\nu}}{P\cdot k_{n+2}}\right) = \frac{1}{t\omega_Pz_{12}}S_P^{(1)}\cr
\lim_{1||2}\mathbb{S}_{12}^{(1,0)} & = &  \frac{i}{(1-t)^2\omega_Pz_{12}}\sum_{b=3}^{n+2}\frac{ \varepsilon\cdot k_b}{P\cdot k_b}  -\frac{i}{(1-t)^2\omega_Pz_{12}}\sum_{b=3}^{n+2}\frac{\varepsilon\cdot k_b}{P\cdot k_b}  = 0 \cr
\lim_{1||2}\mathbb{S}_{12}^{(1,1)} & = & \frac{i}{(1-t)\omega_Pz_{12}}\left(\frac{  \varepsilon_{\rho}P_{\sigma}J_3^{\rho\sigma}}{P\cdot k_3} -\frac{ \varepsilon_{\rho}P_{\sigma}J_{n+2}^{\rho\sigma}}{P\cdot k_{n+2}}\right) = \frac{i}{(1-t)\omega_Pz_{12}}S_P^{(1)}.
\end{eqnarray}
Once again, we see that if we take the sum of all four of these, they organize into the expected soft expansion of the collinear limit, with the correct orders in $\epsilon$ when we take $\epsilon_1=\epsilon_2=\epsilon$. However, we care about the consistency with the soft gluon OPE which was shown in~\cite{Fan:2019emx, Pate:2019lpp} to be given by 
\begin{equation}
R^{k,a}(z_1,\bar{z}_1)R^{l,b}(z_2,\bar{z}_2) \sim \frac{-if^{ab}_c}{z_{12}}\sum_{n=0}^{1-k}\begin{pmatrix} 2-k-l-n\cr 1-l\end{pmatrix}\frac{\bar{z}_{12}^n}{n!}\bar{\partial}^nR^{k+l-1,c}(z_2,\bar{z}_2)
\end{equation}
where $R^{k,a}$ denotes a soft gluon of conformal weight $k\leq 1$ constructed from a celestial operator via a limiting procedure as in the graviton case. This tells us the following
\begin{enumerate}
    \item Two leading soft gluons should give a leading soft gluon. This is consistent with $\lim_{1||2}\mathbb{S}_{12}^{(0,0)}$ and will also be consistent if we switch the ordering of the soft limits. Additionally, what is interesting here, is that since under $t\rightarrow 1-t$, the right hand side is symmetric, the fact that $z_{12}\rightarrow z_{21}$ when we flip the order means the commutator $\mathtt{commSL}^{(0,0)}$ will also give a leading soft gluon while the anticommutator will vanish.
    \item A leading and subleading soft gluon should give a subleading soft gluon. This is consistent with $\mathbb{S}_{12}^{(0,1)}$. It is interesting that $\mathbb{S}_{12}^{(1,0)}$ vanishes but that means that the commutator will also give a subleading soft gluon and be consistent with the OPE.
    \item Two subleading soft gluons should give a sub-subleading soft gluon. We find that we get a subleading soft gluon instead, which is not consistent. This makes sense, based on the argument with derivatives we presented in subsection~\ref{sec:scheme}.
\end{enumerate}
We see that, similar to the gravitational case, we get some consistency until we expect to generate higher order terms in the soft expansion and then it breaks down. 

Recently there has been some discussion about soft algebras in the context of $\mathcal{N}=4$ SYM.~\cite{Alday:2026rso} Since the first two soft theorems are universal in gauge theory, we expect that the above statements hold true for soft gluons in $\mathcal{N}=4$ SYM as well. However, it is very likely that due to the added supersymmetry as well as the ability to resum the full soft expansion, this story becomes more well-defined for this particular gauge theory. We leave the details of this for future work.

\section{Discussion}\label{sec:discussion}
This work began in an effort to understand whether the ``infinite tower of soft operators" was a statement that was true for generic bulk theories of gravity or if it was only valid in the MHV sector where we know soft theorems nicely exponentiate.\cite{Guevara:2019ypd} The goal was to perform a bulk-level computation that connected the results of~\cite{Hamada:2018vrw, Li:2018gnc} with the boundary soft algebras and OPEs. Initially, we wanted to show that if we considered the Cachazo-He-Yuan (CHY) formula~\cite{Cachazo:2013hca} for tree level gravitational amplitudes, we either got consistent relations or that we were unable to use the boundary algebras to generate the soft expansion beyond sub-subleading order unless we specified to the MHV case. However, while performing this computation we realized that the boundary algebras do not give consistent bulk results even in the MHV sector which was quite curious. This led us to concentrate on the first three soft factors, which are universal in gravity, where we found the results presented in this paper. 

We find that if we take the first three soft factors, perform a double soft expansion and then take a collinear limit, the leading terms organize into the first three terms of a soft expansion in the collinear momentum. In particular, this does substantiate the claim that, up to differences in pre-factors, there is a relationship between the two orderings of soft and collinear limits in this context. This also gives us a concrete explanation for why the $\Delta=2$ operator is unphysical~\cite{Guevara:2021abz}. We argue that this is the appropriate bulk interpretation of the boundary celestial OPEs and algebras, which is naturally consistent with how they were originally derived. However, \textbf{the power of the boundary algebras is the statement that the first three soft theorems should be enough to generate all the rest of the terms}, and it is clear from this work, that is not the case for amplitudes in gravity. In particular, none of the expressions derived in this paper involving the sub-subleading soft theorem, appear to relate to any part of the sub$^{(3)}$-leading term in the soft expansion which further solidifies the assertion that we are unable to generate, even a portion of all the terms in a generic soft expansion. 

It would be nice to understand how this fits in with the results of~\cite{Himwich:2023njb}. In their paper they begin with a nice exponentiated result for all the soft factors, written in a nice $SL(2,\mathbb{C})$ covariant way, and reproduce the $w_{1+\infty}$ algebra akin to how it was done originally. It is possible that writing the soft factors as they did, is crucial to making the connection to the algebraic statements. In that case, it would be helpful to understand why using the expressions for the soft factors in terms of angular momentum operators is fundamentally different. We leave this to future work.

Moving foward, we would like to understand what this means for the boundary algebraic statements. It is clear that in the context of celestial CFTs, this infinite dimensional algebra provides a crucial piece of the puzzle, so we would like to understand how that fits in to the bulk physics. We already have nice expressions for the exponentiation of soft factors in the MHV sector~\cite{Guevara:2019ypd} as well as more generic expressions in pure gravity~\cite{Li:2018gnc} and a rudimentary understanding of how this could relate to symmetries~\cite{Hamada:2018vrw} so there should be a way to make those results consistent with the commutation relations we find in the boundary. It is possible that using $\mathcal{N}=4$ SYM as discussed in~\cite{Alday:2026rso} could be a nice playground. Given the results for gauge theory that are presented here, we hope to explore this further in future work. Additionally, it seems like we are able to generate multi-particle contributions in the double soft factors and we would like to explore their connection to the multi-particle contributions to the soft OPEs discussed in~\cite{Guevara:2024ixn}.

\section*{Acknowledgements}
The author would like to thank Freddy Cachazo, Luca Ciambelli and Ana-Maria Raclariu for insightful discussions. In particular, we thank Rajamani Narayanan, Sabrina Pasterski, Andrew Strominger and Tomasz Taylor for comments on early drafts of this work. The author acknowledges support by the Celestial Holography Initiative at the Perimeter Institute for Theoretical Physics and by the Simons Collaboration on Celestial Holography. The author's research at the Perimeter Institute is supported by the Government of Canada through the Department of Innovation, Science and Industry Canada and by the Province of Ontario through the Ministry of Colleges and Universities.

\appendix
\section{Derivatives of soft factors}\label{app:derivatives}
In this appendix we will write general expressions for any number of derivatives acting on the first three soft factors. These will be useful when examining the collinear limits in the main text. Recall that the soft factors are
\begin{eqnarray}
S^{(0)}(1^+) & = & \sum_{a=2}^{n+1}\frac{(\varepsilon_1\cdot k_a)^2}{q_1\cdot k_a}, \ \ S^{(1)}(1^+)  =  -i\sum_{a=2}^{n+1}\frac{\varepsilon_1\cdot k_a}{q_1\cdot k_a}\left(\varepsilon_{1,\nu}q_{1,\rho}J_a^{\rho\nu}\right)\cr
S^{(2)}(1^+) & = & -\frac{1}{2}\sum_{a=2}^{n+1}\frac{\left(\varepsilon_{1,\mu}q_{1,\rho}J_a^{\rho\mu}\right)\left(\varepsilon_{1,\nu}q_{1,\sigma}J_a^{\sigma\nu}\right)}{q_1\cdot k_a}.
\end{eqnarray}
Let's look at some useful derivatives of them. We first notice that 
\begin{eqnarray}
\partial_{z_1}\left(\varepsilon_{1,\nu}q_{1,\rho}J_a^{\rho\nu}\right) & = & \varepsilon_{1,\nu}\varepsilon_{1,\rho}J_a^{\rho\nu}=0, \ \ \partial_{z_1}(\varepsilon_1\cdot k_a)  =  0\cr 
\partial_{z_1}\frac{1}{q_1\cdot k_a}	& = & -\frac{\varepsilon_1\cdot k_a}{(q_1\cdot k_a)^2}, \ \ \partial_{z_1}^\alpha\frac{1}{q_1\cdot k_a} = \frac{(-1)^\alpha \Gamma(\alpha+1)}{(q_1\cdot k_a)^{\alpha+1}}.
\end{eqnarray}
Then we can see
\begin{eqnarray}
\partial_{z_1}^\alpha S^{(0)}(1^+) & = & (-1)^\alpha \Gamma(\alpha+1)\sum_{a=2}^{n+1}\frac{(\varepsilon_1\cdot k_a)^{\alpha+2}}{(q_1\cdot k_a)^{\alpha+1}}\cr
\partial_{z_1}^\alpha S^{(1)}(1^+) & = & -i(-1)^\alpha \Gamma(\alpha+1)\sum_{a=2}^{n+1}\frac{(\varepsilon_1\cdot k_a)^{\alpha+1}}{(q_1\cdot k_a)^{\alpha+1}}\left(\varepsilon_{1,\nu}q_{1,\rho}J_a^{\rho\nu}\right)\cr
\partial_{z_1}^\alpha S^{(2)}(1^+) & = & -\frac{(-1)^\alpha \Gamma(\alpha+1)}{2}\sum_{a=2}^{n+1}\frac{(\varepsilon_1\cdot k_a)^\alpha\left(\varepsilon_{1,\mu}q_{1,\rho}J_a^{\rho\mu}\right)\left(\varepsilon_{1,\nu}q_{1,\sigma}J_a^{\sigma\nu}\right)}{(q_1\cdot k_a)^{\alpha+1}}
\end{eqnarray}

\section{Double soft limits}\label{app:doublesoft}
We will calculate each combination of soft limits at each order. We will order them by epsilon expansion and then take the collinear limit. We first let $S_1^{(m)}S_2^{(n)} \equiv \mathbb{S}_{12}^{(m,n)}$ and note that $\mathbb{S}_{21}^{(m,n)} = \mathbb{S}_{12}^{(m,n)}(1\leftrightarrow 2)$. Therefore, we only need to compute one set of these. We have nine total options. 
\paragraph{$1^+$ Leading and $2^+$ leading:}
\begin{equation}
\mathbb{S}_{12}^{(0,0)} \mathcal{M}_{n+2} = \frac{1}{\epsilon^2}\mathbb{S}_{12}^{(0,0),-2}\mathcal{M}_{n}  + \frac{1}{\epsilon}\mathbb{S}_{12}^{(0,0),-1}\mathcal{M}_{n} 
\end{equation}
where 
\begin{equation}
\mathbb{S}_{12}^{(0,0),-2} = \sum_{a,b=3}^{n+2}\frac{(\varepsilon_1\cdot k_a)^2}{q_1\cdot k_a} \frac{(\varepsilon_2\cdot k_b)^2}{q_2\cdot k_b}, \ \ \mathbb{S}_{12}^{(0,0),-1} = \frac{(\varepsilon_1\cdot q_2)^2}{q_1\cdot q_2} \sum_{a=3}^{n+2}\frac{(\varepsilon_2\cdot k_a)^2}{q_2\cdot k_a}. 
\end{equation}
The collinear limit of this is 
\begin{equation}
\lim_{1||2}\mathbb{S}_{12}^{(0,0),-2} = \frac{1}{t(1-t)}\sum_{a,b=3}^{n+2}\frac{(\varepsilon_P\cdot k_a)^2}{P\cdot k_a} \frac{(\varepsilon_P\cdot k_b)^2}{P\cdot k_b}, \ \ \lim_{1||2}\mathbb{S}_{12}^{(0,0),-1} = -\frac{2}{t}\frac{\bar{z}_{12}}{z_{12}} \sum_{a=3}^{n+2}\frac{(\varepsilon_P\cdot k_a)^2}{P\cdot k_a}. 
\end{equation}
If we switch the particle numbers  $1\leftrightarrow 2$ and take the collinear limit we get
\begin{equation}
\lim_{1||2}\mathbb{S}_{12}^{\ast(0,0),-2} = \frac{1}{t(1-t)}\sum_{a,b=3}^{n+2}\frac{(\varepsilon_P\cdot k_a)^2}{P\cdot k_a} \frac{(\varepsilon_P\cdot k_b)^2}{P\cdot k_b}, \ \ \lim_{1||2}\mathbb{S}_{12}^{\ast(0,0),-1} = -\frac{2}{(1-t)}\frac{\bar{z}_{12}}{z_{12}} \sum_{a=3}^{n+2}\frac{(\varepsilon_P\cdot k_a)^2}{P\cdot k_a}. 
\end{equation}
We can see that the first term is the same, due to the symmetry under particle label exchange, while the second term is not the same.
\paragraph{$1^+$ leading and $2^+$ subleading:}
\begin{equation}
\mathbb{S}_{12}^{(0,1)}\mathcal{M}_{n+2}  = \frac{1}{\epsilon}\mathbb{S}_{12}^{(0,1),-1}\mathcal{M}_{n+2} + \mathbb{S}_{12}^{(0,1),0}\mathcal{M}_{n+2}
\end{equation}
where
\begin{equation}
\mathbb{S}_{12}^{(0,1),-1} = - i\sum_{a,b=3}^{n+2}\frac{(\varepsilon_1\cdot k_a)^2}{q_1\cdot k_a} \frac{(\varepsilon_2\cdot k_b) \left(\varepsilon_{\beta}^2q_{2,\lambda}J_b^{\lambda\beta}\right)}{q_2\cdot k_b}, \ \ \mathbb{S}_{12}^{(0,1),0}= -i\frac{(\varepsilon_1\cdot q_2)^2}{q_1\cdot q_2}  \sum_{a=3}^{n+2}\frac{(\varepsilon_2\cdot k_a) \left(\varepsilon_{\beta}^2 q_{2,\lambda}J_a^{\lambda\beta}\right)}{q_2\cdot k_a}.
\end{equation}
The collinear limit of this is 
\begin{eqnarray}
\lim_{1||2}\mathbb{S}_{12}^{(0,1),-1} & = & - i\sum_{a,b=3}^{n+2}\frac{(\varepsilon_P\cdot k_a)^2}{tP\cdot k_a} \frac{(\varepsilon_P\cdot k_b) \left(\varepsilon_{P,\beta}P_{\lambda}J_b^{\lambda\beta}\right)}{P\cdot k_b}\cr
\lim_{1||2}\mathbb{S}_{12}^{(0,1),0} & = & \frac{2i(1-t)}{t}\frac{\bar{z}_{12}}{z_{12}}  \sum_{a=3}^{n+2}\frac{(\varepsilon_P\cdot k_a) \left(\varepsilon_{P,\beta} P_{\lambda}J_a^{\lambda\beta}\right)}{P\cdot k_a}.
\end{eqnarray}
If we exchange the particle indices and take the collinear limit we get
\begin{eqnarray}
\lim_{1||2}\mathbb{S}_{12}^{\ast(0,1),-1} & = & - \frac{i}{1-t}\sum_{a,b=3}^{n+2}\frac{(\varepsilon_P\cdot k_a)^2}{P\cdot k_a} \frac{(\varepsilon_P\cdot k_b) \left(\varepsilon_{P,\beta}P_{\lambda}J_b^{\lambda\beta}\right)}{P\cdot k_b}\cr
\lim_{1||2}\mathbb{S}_{12}^{\ast(0,1),0} & = & \frac{2it}{(1-t)}  \frac{\bar{z}_{12}}{z_{12}}\sum_{a=3}^{n+2}\frac{(\varepsilon_P\cdot k_a) \left(\varepsilon_{P,\beta} P_{\lambda}J_a^{\lambda\beta}\right)}{P\cdot k_a}.
\end{eqnarray}
\paragraph{$1^+$ leading and $2^+$ sub-subleading:}
\begin{equation}
\mathbb{S}_{12}^{(0,2)}\mathcal{M}_{n+2} = \mathbb{S}_{12}^{(0,2),0}\mathcal{M}_{n+2} + \epsilon\mathbb{S}_{12}^{(0,2),1}\mathcal{M}_{n+2}
\end{equation}
where
\begin{eqnarray}
\mathbb{S}_{12}^{(0,2),0} & = & - \frac{1}{2}\sum_{a,b=3}^{n+2}\frac{(\varepsilon_1\cdot k_a)^2}{q_1\cdot k_a}   \frac{\left(\varepsilon_{\alpha}^2q_{2,\lambda}J_b^{\lambda\alpha}\right)\left(\varepsilon_{\beta}^2q_{2,\tau}J_b^{\tau\beta}\right)}{q_2\cdot k_b}\cr
\mathbb{S}_{12}^{(0,2),1} & = & -\frac{1}{2}\frac{(\varepsilon_1\cdot q_2)^2}{q_1\cdot q_2} \sum_{a=3}^{n+2}\frac{\left(\varepsilon_{\alpha}^2q_{2,\lambda}J_a^{\lambda\alpha}\right)\left(\varepsilon_{\beta}^2q_{2,\tau}J_a^{\tau\beta}\right)}{q_2\cdot k_a}.
\end{eqnarray}
In the collinear limit this is
\begin{eqnarray}
\lim_{1||2}\mathbb{S}_{12}^{(0,2),0} & = & - \frac{1-t}{2t}\sum_{a,b=3}^{n+2}\frac{(\varepsilon_P\cdot k_a)^2}{P\cdot k_a}   \frac{\left(\varepsilon_{P,\alpha}P_{\lambda}J_b^{\lambda\alpha}\right)\left(\varepsilon_{P,\beta}P_{\tau}J_b^{\tau\beta}\right)}{P\cdot k_b}\cr
\lim_{1||2}\mathbb{S}_{12}^{(0,2),1} & = & \frac{(1-t)^2}{t}\frac{\bar{z}_{12}}{z_{12}} \sum_{a=3}^{n+2}\frac{\left(\varepsilon_{P,\alpha}P_{\lambda}J_a^{\lambda\alpha}\right)\left(\varepsilon_{P,\beta}P_{\tau}J_a^{\tau\beta}\right)}{P\cdot k_a}.
\end{eqnarray}
If we exchange the particle indices and take the collinear limit this becomes
\begin{eqnarray}
\lim_{1||2}\mathbb{S}_{12}^{\ast(0,2),0} & = & - \frac{t}{2(1-t)}\sum_{a,b=3}^{n+2}\frac{(\varepsilon_P\cdot k_a)^2}{P\cdot k_a}   \frac{\left(\varepsilon_{P,\alpha}P_{\lambda}J_b^{\lambda\alpha}\right)\left(\varepsilon_{P,\beta}P_{\tau}J_b^{\tau\beta}\right)}{P\cdot k_b}\cr
\lim_{1||2}\mathbb{S}_{12}^{\ast(0,2),1} & = & \frac{t^2}{(1-t)} \frac{\bar{z}_{12}}{z_{12}}\sum_{a=3}^{n+2}\frac{\left(\varepsilon_{P,\alpha}P_{\lambda}J_a^{\lambda\alpha}\right)\left(\varepsilon_{P,\beta}P_{\tau}J_a^{\tau\beta}\right)}{P\cdot k_a}.
\end{eqnarray}
\paragraph{$1^+$ subleading and $2^+$ leading:}
\begin{equation}
\mathbb{S}_{12}^{(1,0)}\mathcal{M}_{n+2} = \frac{1}{\epsilon}\mathbb{S}_{12}^{(1,0),-1}\mathcal{M}_{n+2}
\end{equation}
where 
\begin{eqnarray}
\mathbb{S}_{12}^{(1,0),-1} & = & 2\frac{(\varepsilon_1\cdot q_2)(q_{1}\cdot \varepsilon_2)}{q_1\cdot q_2}\sum_{a=3}^{n+2}\frac{(\varepsilon_2\cdot k_a)(\varepsilon_1\cdot k_a)}{(q_2\cdot k_a)} + \frac{(\varepsilon_1\cdot q_2)^2}{q_1\cdot q_2}\sum_{a=3}^{n+2}\frac{(\varepsilon_2\cdot k_a)^2(q_{1}\cdot k_a)}{(q_2\cdot k_a)^2} \cr
& - & i \sum_{a,b=3}^{n+2}\frac{(\varepsilon_1\cdot k_a) (\varepsilon_2\cdot k_b)^2}{q_1\cdot k_a} \frac{\left(\varepsilon_{\nu}^1 q_{1,\rho}J_a^{\rho\nu}\right)}{q_2\cdot k_b} \cr
& - & 2(\varepsilon_1\cdot q_2)\sum_{a=3}^{n+2}\frac{(\varepsilon_2\cdot k_a)^2(\varepsilon_1\cdot k_a)}{(q_2\cdot k_a)^2} - 2(q_{1}\cdot \varepsilon_2)\sum_{a=3}^{n+2}\frac{(\varepsilon_2\cdot k_a)}{q_2\cdot k_a}   \frac{(\varepsilon_1\cdot k_a)^2  }{q_1\cdot k_a}  \cr
& + & (q_{1}\cdot q_2) \sum_{a=3}^{n+2}\frac{(\varepsilon_1\cdot k_a)^2  }{q_1\cdot k_a} \frac{(\varepsilon_2\cdot k_a)^2 }{(q_2\cdot k_a)^2}.   
\end{eqnarray}
The collinear limit of this is 
\begin{eqnarray}
\lim_{1||2}\mathbb{S}_{12}^{(1,0),-1} & = & \frac{2}{(1-t)}\frac{\bar{z}_{12}}{z_{12}}\sum_{a=3}^{n+2}\frac{(\varepsilon_P\cdot k_a)^2}{(P\cdot k_a)}  - \frac{i}{(1-t)} \sum_{a,b=3}^{n+2}\frac{(\varepsilon_P\cdot k_a) (\varepsilon_P\cdot k_b)^2}{P\cdot k_a} \frac{\left(\varepsilon_{P,\nu} P_{\rho}J_a^{\rho\nu}\right)}{P\cdot k_b} \cr
& - & \frac{2}{1-t}\omega_P^2|z_{12}|^2 \sum_{a=3}^{n+2} \frac{(\varepsilon_P\cdot k_a)^4 }{(P\cdot k_a)^3} .  
\end{eqnarray}
If we exchange the particle indices and take the collinear limit we get
\begin{eqnarray}
\lim_{1||2}\mathbb{S}_{12}^{\ast(1,0),-1} & = & \frac{2}{t}\frac{\bar{z}_{12}}{z_{12}}\sum_{a=3}^{n+2}\frac{(\varepsilon_P\cdot k_a)^2}{(P\cdot k_a)} - \frac{i}{t} \sum_{a,b=3}^{n+2}\frac{(\varepsilon_P\cdot k_a) (\varepsilon_P\cdot k_b)^2}{P\cdot k_a} \frac{\left(\varepsilon_{P,\nu}P_{\rho}J_a^{\rho\nu}\right)}{P\cdot k_b} \cr
& - & \frac{2}{t}\omega_P^2|z_{12}|^2 \sum_{a=3}^{n+2}\frac{(\varepsilon_P\cdot k_a)^4  }{(P\cdot k_a)^3}.  
\end{eqnarray}
\paragraph{$1^+$ subleading and $2^+$ subleading:}
\begin{equation}
\mathbb{S}_{12}^{(1,1)}\mathcal{M}_{n+2} = \mathbb{S}_{12}^{(1,1),0}\mathcal{M}_{n+2}
\end{equation}
where 
\begin{eqnarray}
\mathbb{S}_{12}^{(1,1),0} & = & -\frac{i(\varepsilon_1\cdot q_2) (q_{1}\cdot \varepsilon_2)}{q_1\cdot q_2}  \sum_{a=3}^{n+2}\frac{ \left[(\varepsilon_1\cdot k_a)  \varepsilon_{\beta}^2+(\varepsilon_2\cdot k_a) \varepsilon_{\beta}^1\right] q_{2,\lambda}J_a^{\lambda\beta}}{q_2\cdot k_a}  \cr
& + & \frac{i(\varepsilon_1\cdot q_2)^2 }{q_1\cdot q_2} \sum_{a=3}^{n+2}\frac{(\varepsilon_2\cdot k_a)\left[(q_2\cdot k_a)q_{1,\lambda}-(q_{1}\cdot k_a)q_{2,\lambda}\right]\varepsilon_{\beta}^2J_a^{\lambda\beta}}{(q_2\cdot k_a)^2}\cr
& - &  \sum_{a,b=3}^{n+2}\frac{(\varepsilon_1\cdot k_a) }{q_1\cdot k_a} \frac{(\varepsilon_2\cdot k_b) \left(\varepsilon_{\nu}^1 q_{1,\rho}J_a^{\rho\nu}\right)\left(\varepsilon_{\beta}^2q_{2,\lambda}J_b^{\lambda\beta}\right)}{q_2\cdot k_b} \cr
& - & i(\varepsilon_1\cdot q_2) \sum_{a=3}^{n+2}\frac{(\varepsilon_2\cdot k_a) \left(\varepsilon_{\beta}^2 \varepsilon_{\lambda}^1 J_a^{\lambda\beta}\right)}{q_2\cdot k_a}  +  2i(\varepsilon_1\cdot q_2)  \sum_{a=3}^{n+2}\frac{(\varepsilon_2\cdot k_a)  (\varepsilon_1\cdot k_a) \left(\varepsilon_{\beta}^2 q_{2,\lambda}J_a^{\lambda\beta}\right)}{(q_2\cdot k_a)^2}  \cr
& + &  i(q_{1}\cdot \varepsilon_2)\sum_{a=3}^{n+2}\frac{(\varepsilon_1\cdot k_a)^2 \left(\varepsilon_{\beta}^2q_{2,\lambda}J_a^{\lambda\beta}\right) }{(q_1\cdot k_a)(q_2\cdot k_a)} \cr
& - &  i(q_{1}\cdot q_2)\sum_{a=3}^{n+2}\frac{(\varepsilon_1\cdot k_a)^2 }{q_1\cdot k_a} \frac{(\varepsilon_2\cdot k_a) \left(\varepsilon_{\beta}^2q_{2,\lambda}J_a^{\lambda\beta}\right)}{(q_2\cdot k_a)^2} .
\end{eqnarray}
If we take the collinear limit of this we get 
\begin{eqnarray}
\lim_{1||2}\mathbb{S}_{12}^{(1,1),0} & = & -\frac{4i\bar{z}_{12}}{z_{12}}  \sum_{a=3}^{n+2}\frac{(\varepsilon_P\cdot k_a)  \varepsilon_{P,\beta} P_{\lambda}J_a^{\lambda\beta}}{P\cdot k_a}  - \sum_{a,b=3}^{n+2}\frac{(\varepsilon_P\cdot k_a) }{P\cdot k_a} \frac{(\varepsilon_P\cdot k_b) \left(\varepsilon_{P,\nu} P_{\rho}J_a^{\rho\nu}\right)\left(\varepsilon_{P,\beta}P_{\lambda}J_b^{\lambda\beta}\right)}{P\cdot k_b} \cr
& - &  2i\omega_P\bar{z}_{12}  \sum_{a=3}^{n+2}\frac{(\varepsilon_P\cdot k_a)^2 \left(\varepsilon_{P,\beta}P_{\lambda}J_a^{\lambda\beta}\right)}{(P\cdot k_a)^2}  +  2i\omega_P^2|z_{12}|^2\sum_{a=3}^{n+2} \frac{(\varepsilon_P\cdot k_a)^3 \left(\varepsilon_{P,\beta}P_{\lambda}J_a^{\lambda\beta}\right)}{(P\cdot k_a)^3} 
\end{eqnarray}
If we exchange particle indices and take the collinear limit we get
\begin{eqnarray}
\lim_{1||2}\mathbb{S}_{12}^{\ast(1,1),0} & = & -\frac{4i\bar{z}_{12}}{z_{12}}  \sum_{a=3}^{n+2}\frac{ (\varepsilon_P\cdot k_a)  \varepsilon_{P,\beta}P_{\lambda}J_a^{\lambda\beta}}{P\cdot k_a} -  \sum_{a,b=3}^{n+2}\frac{(\varepsilon_P\cdot k_a) }{P\cdot k_a} \frac{(\varepsilon_P\cdot k_b) \left(\varepsilon_{P,\nu}P_{\rho}J_a^{\rho\nu}\right)\left(\varepsilon_{P,\beta}P_{\lambda}J_b^{\lambda\beta}\right)}{P\cdot k_b} \cr
& + & 2i\omega_P\bar{z}_{12}  \sum_{a=3}^{n+2}\frac{(\varepsilon_P\cdot k_a)^2 \left(\varepsilon_{P,\beta} P_{\lambda}J_a^{\lambda\beta}\right)}{(P\cdot k_a)^2} + 2i\omega_P^2|z_{12}|^2\sum_{a=3}^{n+2} \frac{(\varepsilon_P\cdot k_a)^3 \left(\varepsilon_{P,\beta}P_{\lambda}J_a^{\lambda\beta}\right)}{(P\cdot k_a)^3}. 
\end{eqnarray}
\paragraph{$1^+$ subleading and $2^+$ sub-subleading:}
and 
\begin{equation}
\mathbb{S}_{12}^{(1,2)}\mathcal{M}_{n+2} = \epsilon\mathbb{S}_{12}^{(1,2),1}\mathcal{M}_{n+2}
\end{equation}
where 
\begin{eqnarray}
\mathbb{S}_{12}^{(1,2),1} & = & \frac{1}{2}\frac{(\varepsilon_1\cdot q_2) }{q_1\cdot q_2} \sum_{a=3}^{n+2}\frac{\left[(\varepsilon_1\cdot q_2)\varepsilon_{\alpha}^2q_{1,\lambda}-(\varepsilon_2\cdot q_{1})\varepsilon_{\alpha}^1q_{2,\lambda}\right]\varepsilon_{\beta}^2q_{2,\tau}}{q_2\cdot k_a}\left[J_a^{\lambda\alpha}J_a^{\tau\beta}+J_a^{\tau\beta}J_a^{\lambda\alpha}\right]\cr
& - & \frac{1}{2}\frac{(\varepsilon_1\cdot q_2)^2 }{q_1\cdot q_2}  \sum_{a=3}^{n+2}\frac{  (q_{1}\cdot k_a)}{(q_2\cdot k_a)^2}\left(q_{2,\lambda}\varepsilon_{\alpha}^2J_a^{\lambda\alpha}\right)\left(\varepsilon_{\beta}^2q_{2,\tau} J_a^{\tau\beta}\right)\cr
& + & \frac{i}{2} \sum_{a,b=3}^{n+2}\frac{(\varepsilon_1\cdot k_a) }{q_1\cdot k_a}  \frac{\left(\varepsilon_{\nu}^1q_{1,\rho}J_a^{\rho\nu}\right)\left(\varepsilon_{\alpha}^2q_{2,\lambda}J_b^{\lambda\alpha}\right)\left(q_{2,\tau}\varepsilon_{\beta}^2J_b^{\tau\beta}\right)}{q_2\cdot k_b}\cr
& - & \frac{1}{2}(\varepsilon_1\cdot q_2)  \sum_{a=3}^{n+2}\frac{\varepsilon_{\alpha\beta}^2\varepsilon_{\lambda}^1q_{2,\tau}}{q_2\cdot k_a}\left[J_a^{\lambda\alpha}J_a^{\tau\beta}+J_a^{\tau\beta}J_a^{\lambda\alpha}\right]\cr
& + & (\varepsilon_1\cdot q_2) \sum_{a=3}^{n+2}\frac{(\varepsilon_1\cdot k_a)\left(\varepsilon_{\alpha}^2q_{2,\lambda} J_a^{\lambda\alpha}\right)\left(\varepsilon_{\beta}^2q_{2,\tau}J_a^{\tau\beta}\right)}{(q_2\cdot k_a)^2}\cr
& - & \frac{1}{2}(q_{1}\cdot q_2) \sum_{a=3}^{n+2}\frac{(\varepsilon_1\cdot k_a)^2  }{q_1\cdot k_a}  \frac{\left(\varepsilon_{\alpha}^2q_{2,\lambda}J_a^{\lambda\alpha}\right)\left(\varepsilon_{\beta}^2q_{2,\tau}J_a^{\tau\beta}\right)}{(q_2\cdot k_a)^2}.
\end{eqnarray}
If we take the collinear limit of this we get
\begin{eqnarray}
\lim_{1||2}\mathbb{S}_{12}^{(1,2),1} & = & -3(1-t)\frac{\bar{z}_{12}}{z_{12}} \sum_{a=3}^{n+2}\frac{\varepsilon_{P,\alpha}P_{\lambda}\varepsilon_{P,\beta}P_{\tau}}{P\cdot k_a}J_a^{\lambda\alpha}J_a^{\tau\beta}\cr
& + & \frac{i}{2}(1-t) \sum_{a,b=3}^{n+2}\frac{(\varepsilon_P\cdot k_a) }{P\cdot k_a}  \frac{\left(\varepsilon_{P,\nu}P_{\rho}J_a^{\rho\nu}\right)\left(\varepsilon_{P,\alpha}P_{\lambda}J_b^{\lambda\alpha}\right)\left(P_{\tau}\varepsilon_{P,\beta}J_b^{\tau\beta}\right)}{P\cdot k_b}\cr
& - & 2(1-t)\omega_P\bar{z}_{12} \sum_{a=3}^{n+2}\frac{(\varepsilon_P\cdot k_a)\left(\varepsilon_{P,\alpha}P_{\lambda} J_a^{\lambda\alpha}\right)\left(\varepsilon_{P,\beta}P_{\tau}J_a^{\tau\beta}\right)}{(P\cdot k_a)^2}\cr
& + & (1-t)\omega_P^2|z_{12}|^2 \sum_{a=3}^{n+2}\frac{(\varepsilon_P\cdot k_a)^2  }{(P\cdot k_a)^3}  \left(\varepsilon_{P,\alpha}P_{\lambda}J_a^{\lambda\alpha}\right)\left(\varepsilon_{P,\beta}P_{\tau}J_a^{\tau\beta}\right).
\end{eqnarray}
If we exchange the particle indices and take the collinear limit we get
\begin{eqnarray}
\lim_{1||2}\mathbb{S}_{12}^{\ast(1,2),1} & = & -3t\frac{\bar{z}_{12} }{z_{12}} \sum_{a=3}^{n+2}\frac{\varepsilon_{P,\alpha}P_{\lambda}\varepsilon_{P,\beta}P_{\tau}}{P\cdot k_a}J_a^{\lambda\alpha}J_a^{\tau\beta}\cr
& + & \frac{i}{2}t \sum_{a,b=3}^{n+2}\frac{(\varepsilon_P\cdot k_a) }{P\cdot k_a}  \frac{\left(\varepsilon_{P,\nu}P_{\rho}J_a^{\rho\nu}\right)\left(\varepsilon_{P,\alpha}P_{\lambda}J_b^{\lambda\alpha}\right)\left(P_{\tau}\varepsilon_{P,\beta}J_b^{\tau\beta}\right)}{P\cdot k_b}\cr
& + & 2t\omega_P\bar{z}_{12} \sum_{a=3}^{n+2}\frac{(\varepsilon_P\cdot k_a)\left(\varepsilon_{P,\alpha}P_{\lambda} J_a^{\lambda\alpha}\right)\left(\varepsilon_{P,\beta}P_{\tau}J_a^{\tau\beta}\right)}{(P\cdot k_a)^2}\cr
& + & t\omega_P^2|z_{12}|^2 \sum_{a=3}^{n+2}\frac{(\varepsilon_P\cdot k_a)^2  }{(P\cdot k_a)^3}  \left(\varepsilon_{P,\alpha}P_{\lambda}J_a^{\lambda\alpha}\right)\left(\varepsilon_{P,\beta}P_{\tau}J_a^{\tau\beta}\right).
\end{eqnarray}
\paragraph{$1^+$ sub-subleading and $2^+$ leading:}
\begin{equation}
\mathbb{S}_{12}^{(2,0)}\mathcal{M}_{n+2} = \frac{1}{\epsilon}\mathbb{S}_{12}^{(2,0),-1}\mathcal{M}_{n+2} + \mathbb{S}_{12}^{(2,0),0}\mathcal{M}_{n+2}
\end{equation}
where 
\begin{eqnarray}
\mathbb{S}_{12}^{(2,0),-1} & = & \frac{1}{q_1\cdot q_2}\sum_{a=3}^{n+2}\frac{(q_{1}\cdot k_a)^2}{q_2\cdot k_a}\bigg[ \frac{(q_1\cdot \varepsilon_2)(\varepsilon_1\cdot k_a)}{(q_{1}\cdot k_a)} + \frac{(\varepsilon_1\cdot q_2)(\varepsilon_2\cdot k_a)  }{(q_2\cdot k_a)}\bigg]^2  \cr
& - & 2\sum_{a=3}^{n+2}\frac{(\varepsilon_2\cdot k_a)(\varepsilon_1\cdot k_a)}{(q_2\cdot k_a)^3}\left[(q_{1}\cdot \varepsilon_2)(\varepsilon_1\cdot k_a)   (q_2\cdot k_a)  +(\varepsilon_1\cdot q_2)(\varepsilon_2\cdot k_a)  (q_{1}\cdot k_a)   \right] \cr
& + & (q_{1}\cdot q_2)\sum_{a=3}^{n+2}\frac{(\varepsilon_2\cdot k_a)^2 (\varepsilon_1\cdot k_a)^2}{(q_2\cdot k_a)^3} \cr
\mathbb{S}_{12}^{(2,0),0 }& = &   \sum_{a=3}^{n+2}\frac{q_1\cdot k_a}{q_2\cdot k_a}\left[\frac{(q_{1}\cdot \varepsilon_2)(\varepsilon_1\cdot k_a)}{(q_1\cdot k_a)}+\frac{(\varepsilon_1\cdot q_2)(\varepsilon_2\cdot k_a)}{(q_2\cdot k_a)}\right]^2\cr
& + &  i \sum_{a=3}^{n+2}\frac{(\varepsilon_2\cdot k_a)}{(q_2\cdot k_a)}\left[\frac{2(\varepsilon_1\cdot k_a)(q_{1}\cdot \varepsilon_2)}{(q_1\cdot k_a)}+\frac{(\varepsilon_2\cdot k_a)(\varepsilon_1\cdot q_2)}{(q_2\cdot k_a)}\right]\left(\varepsilon_{\mu}^1q_{1,\rho}J_a^{\rho\mu}\right)\cr
& - &  \frac{1}{2} \sum_{a,b=3}^{n+2}\frac{(\varepsilon_2\cdot k_b)^2}{q_2\cdot k_b}\frac{\left(\varepsilon_{\mu}^1q_{1,\rho}J_a^{\rho\mu}\right)\left(q_{1,\sigma}\varepsilon_{\nu}^1J_a^{\sigma\nu}\right)}{q_1\cdot k_a}\cr
& - &  2(q_{1}\cdot q_2)\sum_{a=3}^{n+2}\frac{(\varepsilon_1\cdot k_a)(\varepsilon_2\cdot k_a)}{(q_2\cdot k_a)^2}\left[\frac{(q_{1}\cdot \varepsilon_2)(\varepsilon_1\cdot k_a) }{(q_1\cdot k_a)}+\frac{(\varepsilon_1\cdot q_2)(\varepsilon_2\cdot k_a)   }{(q_2\cdot k_a)}\right]\cr
& - &  i (q_{1}\cdot q_2)\sum_{a=3}^{n+2}\frac{1}{q_1\cdot k_a}\frac{(\varepsilon_2\cdot k_a)^2 (\varepsilon_1\cdot k_a) }{(q_2\cdot k_a)^2}\left(\varepsilon_{\mu}^1q_{1,\rho}J_a^{\rho\mu}\right)\cr
& + &  (q_{1}\cdot q_2)^2\sum_{a=3}^{n+2}\frac{1}{q_1\cdot k_a}\frac{(\varepsilon_2\cdot k_a)^2(\varepsilon_1\cdot k_a)^2 }{(q_2\cdot k_a)^3}.
\end{eqnarray}
If we take the collinear limit of these we get
\begin{eqnarray}
\mathbb{S}_{12}^{(2,0),-1} & = &  -\frac{2t}{(1-t)^2}\omega_P^2|z_{12}|^2\sum_{a=3}^{n+2}\frac{(\varepsilon_P\cdot k_a)^4}{(P\cdot k_a)^3} \cr
\mathbb{S}_{12}^{(2,0),0 }& = &  \frac{2it}{1-t} \omega_P\bar{z}_{12}\sum_{a=3}^{n+2}\frac{(\varepsilon_P\cdot k_a)^2}{(P\cdot k_a)^2}\left(\varepsilon_{P,\mu}P_{\rho}J_a^{\rho\mu}\right) - \frac{1}{2} \frac{t}{1-t}\sum_{a,b=3}^{n+2}\frac{(\varepsilon_P\cdot k_b)^2}{P\cdot k_b}\frac{\left(\varepsilon_{P,\mu}P_{\rho}J_a^{\rho\mu}\right)\left(\varepsilon_{P,\nu}P_{\sigma}J_a^{\sigma\nu}\right)}{P\cdot k_a}\cr
& + & \frac{2it}{(1-t)}\omega_P^2|z_{12}|^2\sum_{a=3}^{n+2}\frac{(\varepsilon_P\cdot k_a)^3}{(P\cdot k_a)^3}\left(\varepsilon_{P,\mu}P_{\rho}J_a^{\rho\mu}\right) + \frac{4t}{(1-t)}\omega_P^4|z_{12}|^4\sum_{a=3}^{n+2}\frac{(\varepsilon_P\cdot k_a)^4}{(P\cdot k_a)^4}
\end{eqnarray}
If we exchange the particle indices and take the collinear limit we obtain
\begin{eqnarray}
\lim_{1||2}\mathbb{S}_{12}^{\ast(2,0),-1} & = & -\frac{2(1-t)}{t^2}\omega_P^2|z_{12}|^2\sum_{a=3}^{n+2}\frac{(\varepsilon_P\cdot k_a)^4}{(P\cdot k_a)^3} \cr
\lim_{1||2}\mathbb{S}_{12}^{\ast(2,0),0 }& = &  -\frac{2i(1-t)}{t}\omega_P\bar{z}_{12} \sum_{a=3}^{n+2}\frac{(\varepsilon_P\cdot k_a)^2}{(P\cdot k_a)^2}\left(\varepsilon_{P,\mu}P_{\rho}J_a^{\rho\mu}\right)\cr
& - &  \frac{1}{2}\frac{1-t}{t} \sum_{a,b=3}^{n+2}\frac{(\varepsilon_P\cdot k_b)^2}{P\cdot k_b}\frac{\left(\varepsilon_{P,\mu}P_{\rho}J_a^{\rho\mu}\right)\left(P_{\sigma}\varepsilon_{P,\nu}J_a^{\sigma\nu}\right)}{P\cdot k_a}\cr
& + & \frac{2i(1-t)}{t}\omega_P^2|z_{12}|^2\sum_{a=3}^{n+2}\frac{(\varepsilon_P\cdot k_a)^3}{(P\cdot k_a)^3}\left(\varepsilon_{P,\mu}P_{\rho}J_a^{\rho\mu}\right) +  \frac{4(1-t)}{t}\omega_P^4|z_{12}|^4\sum_{a=3}^{n+2}\frac{(\varepsilon_P\cdot k_a)^4 }{(P\cdot k_a)^4}
\end{eqnarray}
\paragraph{$1^+$ sub-subleading and $2^+$ subleading:}
\begin{equation}
\mathbb{S}_{12}^{(2,1)}\mathcal{M}_{n+2} = \mathbb{S}_{12}^{(2,1),0}\mathcal{M}_{n+2} + \epsilon\mathbb{S}_{12}^{(2,1),1}\mathcal{M}_{n+2}
\end{equation}
where 
\begin{eqnarray}
\mathbb{S}_{12}^{(2,1),0} & = &i\frac{1}{q_1\cdot q_2}\sum_{a=3}^{n+2}\frac{\left[(\varepsilon_1\cdot q_2)(\varepsilon_2\cdot k_a)   (q_{1}\cdot k_a) +(q_{1}\cdot \varepsilon_2) (\varepsilon_1\cdot k_a)(q_2\cdot k_a) \right]}{(q_2\cdot k_a)^2}\cr
& \times & \left[(\varepsilon_1\cdot q_2)\varepsilon_{\beta}^2q_{1,\lambda}-(q_{1}\cdot \varepsilon_2) \varepsilon_{\beta}^1q_{2,\lambda}\right]J_a^{\lambda\beta} \cr
& - & i\frac{(\varepsilon_1\cdot q_2)}{q_1\cdot q_2}\sum_{a=3}^{n+2}\frac{(q_{1}\cdot k_a)\left[(\varepsilon_1\cdot q_2)(\varepsilon_2\cdot k_a) (q_{1}\cdot k_a) +(q_{1}\cdot \varepsilon_2) (\varepsilon_1\cdot k_a)(q_2\cdot k_a)\right]}{(q_2\cdot k_a)^3}\left(\varepsilon_{\beta}^2q_{2,\lambda}J_a^{\lambda\beta}\right)  \cr
& + & i\frac{(q_{1}\cdot \varepsilon_2) (\varepsilon_1\cdot q_2)}{q_1\cdot q_2}\sum_{a=3}^{n+2}\frac{(\varepsilon_2\cdot k_a) \left( \varepsilon_{\beta}^1q_{1,\lambda}J_a^{\lambda\beta}\right)}{q_2\cdot k_a}  \cr
& - & i\sum_{a=3}^{n+2}\frac{\left[(q_{1}\cdot \varepsilon_2) (\varepsilon_1\cdot k_a)(q_2\cdot k_a) +(\varepsilon_1\cdot q_2)(\varepsilon_2\cdot k_a)  (q_{1}\cdot k_a)\right]}{(q_2\cdot k_a)^2} \left(\varepsilon_{\lambda}^1 \varepsilon_{\beta}^2J_a^{\lambda\beta}\right) \cr
& + & i\sum_{a=3}^{n+2}\frac{(\varepsilon_2\cdot k_a)(\varepsilon_1\cdot k_a)\left[(q_{1}\cdot \varepsilon_2)    \varepsilon_{\beta}^1q_{2,\lambda}-(\varepsilon_1\cdot q_2)   \varepsilon_{\beta}^2q_{1,\lambda}\right]}{(q_2\cdot k_a)^2}J_a^{\lambda\beta}  \cr
& + & i\sum_{a=3}^{n+2}\frac{(\varepsilon_1\cdot k_a)}{(q_2\cdot k_a)^3}\left[(q_{1}\cdot \varepsilon_2) (\varepsilon_1\cdot k_a)(q_2\cdot k_a) +2(\varepsilon_1\cdot q_2)(\varepsilon_2\cdot k_a)     (q_{1}\cdot k_a) \right] \left(\varepsilon_{\beta}^2q_{2,\lambda}J_a^{\lambda\beta}\right) \cr
& + & i(q_{1}\cdot q_2)\sum_{a=3}^{n+2}\frac{(\varepsilon_2\cdot k_a)(\varepsilon_1\cdot k_a) \left(\varepsilon_{\beta}^2 \varepsilon_{\lambda}^1J_a^{\lambda\beta}\right)}{(q_2\cdot k_a)^2} \cr
& - & i(q_{1}\cdot q_2)\sum_{a=3}^{n+2}\frac{(\varepsilon_2\cdot k_a) (\varepsilon_1\cdot k_a)^2 \left(\varepsilon_{\beta}^2q_{2,\lambda}J_a^{\lambda\beta}\right)}{(q_2\cdot k_a)^3}\cr
\mathbb{S}_{12}^{(2,1),1} & = &  \frac{i}{2}\sum_{a,b=3}^{n+2}\frac{1}{q_1\cdot k_a} \frac{(\varepsilon_2\cdot k_b) \left(\varepsilon_{\mu}^1q_{1,\rho}J_a^{\rho\mu}\right)\left(\varepsilon_{\nu}^1q_{1,\sigma}J_a^{\sigma\nu}\right)\left(\varepsilon_{\beta}^2q_{2,\lambda}J_b^{\lambda\beta}\right)}{q_2\cdot k_b} \cr
& + & \sum_{a=3}^{n+2}\frac{1}{q_2\cdot k_a}\left[\frac{(q_{1}\cdot \varepsilon_2)(\varepsilon_1\cdot k_a)}{q_1\cdot k_a}   + \frac{  (\varepsilon_1\cdot q_2)(\varepsilon_2\cdot k_a) }{q_2\cdot k_a}\right] \left(\varepsilon_{\mu}^1q_{1,\rho}J_a^{\rho\mu}\right)\left(\varepsilon_{\beta}^2q_{2,\lambda}J_a^{\lambda\beta}\right)\cr
& - & i(\varepsilon_1\cdot q_2)\sum_{a=3}^{n+2}\frac{\left[ (\varepsilon_1\cdot q_2) (q_{1}\cdot k_a) (\varepsilon_2\cdot k_a) + (q_{1}\cdot \varepsilon_2)(\varepsilon_1\cdot k_a) (q_2\cdot k_a)\right]}{(q_2\cdot k_a)^3}\left(\varepsilon_{\beta}^2q_{2,\lambda}J_a^{\lambda\beta}\right)\cr
& - & (q_{1}\cdot q_2)\sum_{a=3}^{n+2}\frac{1}{q_1\cdot k_a} \frac{(\varepsilon_1\cdot k_a)(\varepsilon_2\cdot k_a) \left(\varepsilon_{\nu}^1q_{1,\sigma}J_a^{\sigma\nu}\right)\left(\varepsilon_{\beta}^2q_{2,\lambda}J_a^{\lambda\beta}\right)}{(q_2\cdot k_a)^2} \cr
& + &  i(q_{1}\cdot q_2)\sum_{a=3}^{n+2}\frac{(\varepsilon_1\cdot k_a)}{(q_2\cdot k_a)^2}\left[\frac{(q_{1}\cdot \varepsilon_2) (\varepsilon_1\cdot k_a)}{q_1\cdot k_a}   +\frac{ 2(\varepsilon_1\cdot q_2) (\varepsilon_2\cdot k_a) }{q_2\cdot k_a}\right] \left(\varepsilon_{\beta}^2q_{2,\lambda}J_a^{\lambda\beta}\right)\cr
& - & i(q_{1}\cdot q_2)^2\sum_{a=3}^{n+2}\frac{1}{q_1\cdot k_a} \frac{(\varepsilon_1\cdot k_a)^2 (\varepsilon_2\cdot k_a) \left(\varepsilon_{\beta}^2q_{2,\lambda}J_a^{\lambda\beta}\right)}{(q_2\cdot k_a)^3} 
\end{eqnarray}
If we take the collinear limit of this we get
\begin{eqnarray}
\mathbb{S}_{12}^{(2,1),0} & = & \frac{2it}{1-t}\frac{\bar{z}_{12}}{z_{12}}\sum_{a=3}^{n+2}\frac{(\varepsilon_P\cdot k_a) \left( \varepsilon_{P,\beta}P_{\lambda}J_a^{\lambda\beta}\right)}{P\cdot k_a} + \frac{2it}{1-t}\omega_P\bar{z}_{12}\sum_{a=3}^{n+2}\frac{(\varepsilon_P\cdot k_a)^2    \varepsilon_{P,\beta}P_{\lambda}}{(P\cdot k_a)^2}J_a^{\lambda\beta}  \cr
& + & \frac{2it}{1-t}\omega_P^2|z_{12}|^2\sum_{a=3}^{n+2}\frac{(\varepsilon_P\cdot k_a)^3 \left(\varepsilon_{P,\beta}P_{\lambda}J_a^{\lambda\beta}\right)}{(P\cdot k_a)^3}\cr
\mathbb{S}_{12}^{(2,1),1} & = &  \frac{it}{2}\sum_{a,b=3}^{n+2}\frac{1}{P\cdot k_a} \frac{(\varepsilon_P\cdot k_b) \left(\varepsilon_{P,\mu}P_{\rho}J_a^{\rho\mu}\right)\left(\varepsilon_{P,\nu}P_{\sigma}J_a^{\sigma\nu}\right)\left(\varepsilon_{P,\beta}P_{\lambda}J_b^{\lambda\beta}\right)}{P\cdot k_b} \cr
& + & 2t\omega_P^2|z_{12}|^2\sum_{a=3}^{n+2}\frac{(\varepsilon_P\cdot k_a)^2 \left(\varepsilon_{P,\nu}P_{\sigma}J_a^{\sigma\nu}\right)\left(\varepsilon_{P,\beta}P_{\lambda}J_a^{\lambda\beta}\right)}{(P\cdot k_a)^3} \cr
& + &  4it\omega_P^3|z_{12}|^2\bar{z}_{12}\sum_{a=3}^{n+2}\frac{(\varepsilon_P\cdot k_a)^2}{(P\cdot k_a)^3} \left(\varepsilon_{P,\beta}P_{\lambda}J_a^{\lambda\beta}\right) - 4it\omega_P^4|z_{12}|^4\sum_{a=3}^{n+2} \frac{(\varepsilon_P\cdot k_a)^3 \left(\varepsilon_{P,\beta}P_{\lambda}J_a^{\lambda\beta}\right)}{(P\cdot k_a)^4} 
\end{eqnarray}
If we exchange particle indices and take the collinear limit we get
\begin{eqnarray}
\lim_{1||2} \mathbb{S}_{12}^{\ast(2,1),0} & = & \frac{2i(1-t)}{t}\frac{\bar{z}_{12}}{z_{12}}\sum_{a=3}^{n+2}\frac{(\varepsilon_P\cdot k_a) \left( \varepsilon_{P,\beta}P_{\lambda}J_a^{\lambda\beta}\right)}{P\cdot k_a} - \frac{2i(1-t)}{t}\omega_P\bar{z}_{12}\sum_{a=3}^{n+2}\frac{(\varepsilon_P\cdot k_a)^2}{(P\cdot k_a)^2}\left(\varepsilon_{P,\beta}P_{\lambda}J_a^{\lambda\beta}\right) \cr
& + & \frac{2i(1-t)}{t}\omega_P^2|z_{12}|^2\sum_{a=3}^{n+2}\frac{ (\varepsilon_P\cdot k_a)^3 \left(\varepsilon_{P,\beta}P_{\lambda}J_a^{\lambda\beta}\right)}{(P\cdot k_a)^3}\cr
\lim_{1||2}\mathbb{S}_{12}^{\ast(2,1),1} & = &  \frac{i(1-t)}{2}\sum_{a,b=3}^{n+2}\frac{1}{P\cdot k_a} \frac{(\varepsilon_P\cdot k_b) \left(\varepsilon_{P,\mu}P_{\rho}J_a^{\rho\mu}\right)\left(\varepsilon_{P,\nu}P_{\sigma}J_a^{\sigma\nu}\right)\left(\varepsilon_{P,\beta}P_{\lambda}J_b^{\lambda\beta}\right)}{P\cdot k_b} \cr
& + & 2(1-t)\omega_P^2|z_{12}|^2\sum_{a=3}^{n+2} \frac{(\varepsilon_P\cdot k_a)^2 \left(\varepsilon_{P,\nu}P_{\sigma}J_a^{\sigma\nu}\right)\left(\varepsilon_{P,\beta}P_{\lambda}J_a^{\lambda\beta}\right)}{(P\cdot k_a)^3} \cr
& - &  4i(1-t)\omega_P^3|z_{12}|^2\bar{z}_{12}\sum_{a=3}^{n+2}\frac{(\varepsilon_P\cdot k_a)}{(P\cdot k_a)^2}\frac{(\varepsilon_P\cdot k_a) }{P\cdot k_a} \left(\varepsilon_{P,\beta}P_{\lambda}J_a^{\lambda\beta}\right)\cr
& - & 4i(1-t)\omega_P^4|z_{12}|^4\sum_{a=3}^{n+2}\frac{1}{P\cdot k_a} \frac{(\varepsilon_P\cdot k_a)^3 \left(\varepsilon_{P,\beta}P_{\lambda}J_a^{\lambda\beta}\right)}{(P\cdot k_a)^3} 
\end{eqnarray}
\paragraph{$1^+$ sub-subleading and $2^+$ sub-subleading:}
\begin{equation}
\mathbb{S}_{12}^{(2,2)}\mathcal{M}_{n+2} = \epsilon\mathbb{S}_{12}^{(2,2),1}\mathcal{M}_{n+2} + \epsilon^2\mathbb{S}_{12}^{(2,2),2}\mathcal{M}_{n+2}
\end{equation}
\begin{eqnarray}
\mathbb{S}_{12}^{(2,2),1} & = &-\frac{1}{2}\frac{(q_{1}\cdot \varepsilon_2)^2}{q_1\cdot q_2}\sum_{a=3}^{n+2}\frac{1}{q_2\cdot k_a}\left(\varepsilon_{\alpha}^1q_{2,\lambda}J_a^{\lambda\alpha}\right)\left(\varepsilon_{\beta}^1q_{2,\tau}J_a^{\tau\beta}\right)\cr
& - & \frac{1}{2}\frac{(\varepsilon_1\cdot q_2)^2}{q_1\cdot q_2}\sum_{a=3}^{n+2}\frac{1}{q_2\cdot k_a}\left(\varepsilon_{\alpha}^2q_{1,\lambda}J_a^{\lambda\alpha}\right)\left(\varepsilon_{\beta}^2q_{1,\tau}J_a^{\tau\beta}\right)\cr
& + & \frac{1}{2}\frac{(q_{1}\cdot \varepsilon_2) (\varepsilon_1\cdot q_2)}{q_1\cdot q_2}\sum_{a=3}^{n+2}\frac{\varepsilon_{\alpha}^1\varepsilon_{\beta}^2(q_{1,\tau}q_{2,\lambda}+q_{1,\lambda}q_{2,\tau})}{q_2\cdot k_a}\left[J_a^{\lambda\alpha}J_a^{\tau\beta}+J_a^{\tau\beta}J_a^{\lambda\alpha}\right]\cr
& - & \frac{1}{2}\frac{(q_{1}\cdot \varepsilon_2) (\varepsilon_1\cdot q_2)}{q_1\cdot q_2}\sum_{a=3}^{n+2}\frac{ (q_{1}\cdot k_a)\varepsilon_{\alpha}^1\varepsilon_{\beta}^2q_{2,\lambda}q_{2,\tau}}{(q_2\cdot k_a)^2}\left[J_a^{\lambda\alpha}J_a^{\tau\beta}+J_a^{\tau\beta}J_a^{\lambda\alpha}\right]\cr
& + & \frac{1}{2}\frac{(\varepsilon_1\cdot q_2)^2}{q_1\cdot q_2}\sum_{a=3}^{n+2}\frac{(q_{1}\cdot k_a)\varepsilon_{\alpha}^2\varepsilon_{\beta}^2q_{1,\lambda}q_{2,\tau}}{(q_2\cdot k_a)^2}\left[J_a^{\lambda\alpha}J_a^{\tau\beta}+J_a^{\tau\beta}J_a^{\lambda\alpha}\right]\cr
& - & \frac{1}{2}\frac{(\varepsilon_1\cdot q_2)^2}{q_1\cdot q_2}\sum_{a=3}^{n+2}\frac{(q_{1}\cdot k_a)^2 }{(q_2\cdot k_a)^3}\left(\varepsilon_{\alpha}^2q_{2,\lambda}J_a^{\lambda\alpha}\right)\left(\varepsilon_{\beta}^2q_{2,\tau}J_a^{\tau\beta}\right)\cr
& - & \frac{1}{2}(q_{1}\cdot \varepsilon_2)\sum_{a=3}^{n+2}\frac{\varepsilon_{\alpha}^1\varepsilon_{\tau}^1\varepsilon_{\beta}^2q_{2,\lambda}}{q_2\cdot k_a}\left[J_a^{\lambda\alpha}J_a^{\tau\beta}+J_a^{\tau\beta}J_a^{\lambda\alpha}\right]\cr
& + & \frac{1}{2}(\varepsilon_1\cdot q_2)\sum_{a=3}^{n+2}\frac{\varepsilon_{\lambda}^1\varepsilon_{\alpha}^2\varepsilon_{\beta}^2q_{1,\tau}}{q_2\cdot k_a}\left[J_a^{\lambda\alpha}J_a^{\tau\beta}+J_a^{\tau\beta}J_a^{\lambda\alpha}\right]\cr
& - & \frac{1}{2}(\varepsilon_1\cdot q_2)\sum_{a=3}^{n+2}\frac{ (q_{1}\cdot k_a)\varepsilon_{\lambda}^1\varepsilon_{\alpha}^2\varepsilon_{\beta}^2q_{2,\tau}}{(q_2\cdot k_a)^2}\left[J_a^{\lambda\alpha}J_a^{\tau\beta}+J_a^{\tau\beta}J_a^{\lambda\alpha}\right]\cr
& + & \frac{1}{2}(q_{1}\cdot \varepsilon_2)\sum_{a=3}^{n+2}\frac{ (\varepsilon_1\cdot k_a)\varepsilon_{\alpha}^1\varepsilon_{\beta}^2q_{2,\tau}q_{2,\lambda}}{(q_2\cdot k_a)^2}\left[J_a^{\lambda\alpha}J_a^{\tau\beta}+J_a^{\tau\beta}J_a^{\lambda\alpha}\right]\cr
& - & \frac{1}{2}(\varepsilon_1\cdot q_2)\sum_{a=3}^{n+2}\frac{  (\varepsilon_1\cdot k_a)\varepsilon_{\alpha}^2\varepsilon_{\beta}^2q_{1,\lambda}q_{2,\tau}}{(q_2\cdot k_a)^2}\left[J_a^{\lambda\alpha}J_a^{\tau\beta}+J_a^{\tau\beta}J_a^{\lambda\alpha}\right]\cr
& + & (\varepsilon_1\cdot  q_2)\sum_{a=3}^{n+2}\frac{(\varepsilon_1\cdot k_a) (q_{1}\cdot k_a)}{(q_2\cdot k_a)^3}\left(q_{2,\lambda}\varepsilon_{\alpha}^2J_a^{\lambda\alpha}\right)\left(q_{2,\tau}\varepsilon_{\beta}^2J_a^{\tau\beta}\right)\cr
& - & \frac{1}{2}(q_{1}\cdot q_2)\sum_{a=3}^{n+2}\frac{1}{q_2\cdot k_a}\left(\varepsilon_{\lambda}^1\varepsilon_{\alpha}^2J_a^{\lambda\alpha}\right)\left(\varepsilon_{\tau}^1\varepsilon_{\beta}^2J_a^{\tau\beta}\right)\cr
& - & \frac{1}{2}(q_{1}\cdot q_2)\sum_{a=3}^{n+2}\frac{(\varepsilon_1\cdot k_a)^2}{(q_2\cdot k_a)^3}\left(q_{2,\lambda}\varepsilon_{\alpha}^2J_a^{\lambda\alpha}\right)\left(q_{2,\tau}\varepsilon_{\beta}^2J_a^{\tau\beta}\right)\cr
& + & \frac{1}{2}(q_{1}\cdot q_2)\sum_{a=3}^{n+2}\frac{ (\varepsilon_1\cdot k_a)\varepsilon_{\lambda}^1\varepsilon_{\alpha}^2q_{2,\tau}\varepsilon_{\beta}^2}{(q_2\cdot k_a)^2}\left[J_a^{\lambda\alpha}J_a^{\tau\beta}+J_a^{\tau\beta}J_a^{\lambda\alpha}\right]\cr
\mathbb{S}_{12}^{(2,2),2} & = & \frac{1}{4} \sum_{a,b=3}^{n+2}\frac{1}{q_1\cdot k_a} \frac{\left(q_{1,\rho}\varepsilon_{\mu}^1J_a^{\rho\mu}\right)\left(q_{1,\sigma}\varepsilon_{\nu}^1J_a^{\sigma\nu}\right)\left(q_{2,\lambda}\varepsilon_{\alpha}^2J_b^{\lambda\alpha}\right)\left(q_{2,\tau}\varepsilon_{\beta}^2J_b^{\tau\beta}\right)}{q_2\cdot k_b} \cr
& - & \frac{i}{2} (\varepsilon_1\cdot q_2)\sum_{a=3}^{n+2} \frac{ \left(q_{1,\sigma}\varepsilon_{\nu}^1J_a^{\sigma\nu}\right)\left(q_{2,\lambda}\varepsilon_{\alpha}^2J_a^{\lambda\alpha}\right)\left(q_{2,\tau}\varepsilon_{\beta}^2J_a^{\tau\beta}\right)}{(q_2\cdot k_a)^2} \cr
& + & \frac{i}{2} (q_{1}\cdot q_2)\sum_{a=3}^{n+2}\frac{1}{q_1\cdot k_a} \frac{(\varepsilon_1\cdot k_a)\left(q_{1,\sigma}\varepsilon_{\nu}^1J_a^{\sigma\nu}\right)\left(q_{2,\lambda}\varepsilon_{\alpha}^2J_a^{\lambda\alpha}\right)\left(q_{2,\tau}\varepsilon_{\beta}^2J_a^{\tau\beta}\right)}{(q_2\cdot k_a)^2} \cr
& - & \frac{1}{2} \sum_{a=3}^{n+2}\frac{\left[(\varepsilon_1\cdot q_2)(q_{1}\cdot k_a)    -(q_{1}\cdot q_2)(\varepsilon_1\cdot k_a)\right]^2}{(q_2\cdot k_a)^3(q_1\cdot k_a)}\left(q_{2,\lambda}\varepsilon_{\alpha}^2J_a^{\lambda\alpha}\right)  \left(q_{2,\tau}\varepsilon_{\beta}^2J_a^{\tau\beta}\right)
\end{eqnarray}
If we take the collinear limit of this we get
\begin{eqnarray}
\mathbb{S}_{12}^{(2,2),1} & = & 3t\frac{\bar{z}_{12}}{z_{12}}\sum_{a=3}^{n+2}\frac{1}{P\cdot k_a}\left(\varepsilon_{P,\alpha}P_{\lambda}J_a^{\lambda\alpha}\right)\left(\varepsilon_{P,\beta}P_{\tau}J_a^{\tau\beta}\right) \cr
& + & 2t\omega_P\bar{z}_{12}\sum_{a=3}^{n+2}\frac{ (\varepsilon_P\cdot k_a)}{(P\cdot k_a)^2}\left(\varepsilon_{P,\alpha}P_{\lambda}J_a^{\lambda\alpha}\right)\left(\varepsilon_{P,\beta}P_{\tau}J_a^{\tau\beta}\right)\cr
& + & t\omega_P^2|z_{12}|^2\sum_{a=3}^{n+2}\frac{(\varepsilon_P\cdot k_a)^2}{(P\cdot k_a)^3}\left(\varepsilon_{P,\alpha}^2P_{\lambda}J_a^{\lambda\alpha}\right)\left(\varepsilon_{P,\beta}^2P_{\tau}J_a^{\tau\beta}\right)\cr
\mathbb{S}_{12}^{(2,2),2} & = & \frac{t(1-t)}{4} \sum_{a,b=3}^{n+2}\frac{1}{P\cdot k_a} \frac{\left(P_{\rho}\varepsilon_{P,\mu}J_a^{\rho\mu}\right)\left(P_{\sigma}\varepsilon_{P,\nu}J_a^{\sigma\nu}\right)\left(P_{\lambda}\varepsilon_{P,\alpha}J_b^{\lambda\alpha}\right)\left(P_{\tau}\varepsilon_{P,\beta}J_b^{\tau\beta}\right)}{P\cdot k_b} \cr
& + & i t(1-t)\omega_P\bar{z}_{12}\sum_{a=3}^{n+2} \frac{ \left(P_{\sigma}\varepsilon_{P,\nu}J_a^{\sigma\nu}\right)\left(P_{\lambda}\varepsilon_{P,\alpha}J_a^{\lambda\alpha}\right)\left(P_{\tau}\varepsilon_{P,\beta}J_a^{\tau\beta}\right)}{(P\cdot k_a)^2} \cr
& - & it(1-t)\omega_P^2|z_{12}|^2\sum_{a=3}^{n+2} \frac{(\varepsilon_P\cdot k_a)\left(P_{\sigma}\varepsilon_{P,\nu}J_a^{\sigma\nu}\right)\left(P_{\lambda}\varepsilon_{P,\alpha}J_a^{\lambda\alpha}\right)\left(P_{\tau}\varepsilon_{P,\beta}J_a^{\tau\beta}\right)}{(P\cdot k_a)^3} \cr
& - & 2t(1-t)\omega_P^2\bar{z}_{12}^2 \sum_{a=3}^{n+2}\frac{\left[(P\cdot k_a) -\omega_Pz_{12}(\varepsilon_P\cdot k_a)\right]^2}{(P\cdot k_a)^4}\left(P_{\lambda}\varepsilon_{P,\alpha}J_a^{\lambda\alpha}\right)  \left(P_{\tau}\varepsilon_{P,\beta}J_a^{\tau\beta}\right)
\end{eqnarray}
If we exchange the particle indices and take the collinear limit we obtain
\begin{eqnarray}
\mathbb{S}_{12}^{\ast(2,2),1} & = & 3(1-t)\frac{\bar{z}_{12}}{z_{12}}\sum_{a=3}^{n+2}\frac{1}{P\cdot k_a}\left(\varepsilon_{P,\alpha}P_{\lambda}J_a^{\lambda\alpha}\right)\left(\varepsilon_{P,\beta}P_{\tau}J_a^{\tau\beta}\right)\cr
& - & 2(1-t)\omega_P\bar{z}_{12}\sum_{a=3}^{n+2}\frac{ (\varepsilon_P\cdot k_a)}{(P\cdot k_a)^2}\left(\varepsilon_{P,\alpha}P_{\lambda}J_a^{\lambda\alpha}\right)\left(\varepsilon_{P,\beta}P_{\tau}J_a^{\tau\beta}\right)\cr
& + & (1-t)\omega_P^2|z_{12}|^2\sum_{a=3}^{n+2}\frac{(\varepsilon_P\cdot k_a)^2}{(P\cdot k_a)^3}\left(P_{\lambda}\varepsilon_{P,\alpha}J_a^{\lambda\alpha}\right)\left(P_{\tau}\varepsilon_{P,\beta}J_a^{\tau\beta}\right)\cr
\lim_{1||2}\mathbb{S}_{12}^{\ast(2,2),2} & = & \frac{t(1-t)}{4} \sum_{a,b=3}^{n+2}\frac{1}{P\cdot k_a} \frac{\left(P_{\rho}\varepsilon_{P,\mu}J_a^{\rho\mu}\right)\left(P_{\sigma}\varepsilon_{P,\nu}J_a^{\sigma\nu}\right)\left(P_{\lambda}\varepsilon_{P,\alpha}J_b^{\lambda\alpha}\right)\left(P_{\tau}\varepsilon_{P,\beta}J_b^{\tau\beta}\right)}{P\cdot k_b} \cr
& - & it(1-t)\omega_P\bar{z}_{12}\sum_{a=3}^{n+2} \frac{ \left(P_{\sigma}\varepsilon_{P,\nu}J_a^{\sigma\nu}\right)\left(P_{\lambda}\varepsilon_{P,\alpha}J_a^{\lambda\alpha}\right)\left(P_{\tau}\varepsilon_{P,\beta}J_a^{\tau\beta}\right)}{(P\cdot k_a)^2} \cr
& - & it(1-t)\omega_P^2|z_{12}|^2\sum_{a=3}^{n+2} \frac{(\varepsilon_P\cdot k_a)\left(P_{\sigma}\varepsilon_{P,\nu}J_a^{\sigma\nu}\right)\left(P_{\lambda}\varepsilon_{P,\alpha}J_a^{\lambda\alpha}\right)\left(P_{\tau}\varepsilon_{P,\beta}J_a^{\tau\beta}\right)}{(P\cdot k_a)^3} \cr
& - & 2t(1-t)\omega_P^2\bar{z}_{12}^2 \sum_{a=3}^{n+2}\frac{\left[(P\cdot k_a) + \omega_Pz_{12}(\varepsilon_P\cdot k_a)\right]^2}{(P\cdot k_a)^4}\left(P_{\lambda}\varepsilon_{P,\alpha}J_a^{\lambda\alpha}\right)  \left(P_{\tau}\varepsilon_{P,\beta}J_a^{\tau\beta}\right)
\end{eqnarray}

\bibliographystyle{utphys}
\bibliography{cpb}

\end{document}